\documentstyle[11pt]{article}

\begin{document}

\newcommand{\beq}{\begin{equation}}
\newcommand{\eeq}{\end{equation}}
\newcommand{\bea}{\begin{eqnarray}}
\newcommand{\eea}{\end{eqnarray}}

\newcommand{\chii}{\raise.5ex\hbox{$\chi$}}
\newcommand{\Z}{Z \! \! \! Z}
\newcommand{\R}{I \! \! R}
\newcommand{\N}{I \! \! N}
\newcommand{\C}{I \! \! \! \! C}

\newcommand{\noi}{\noindent}
\newcommand{\vs}{\vspace{5mm}}
\newcommand{\ie}{{${ i.e.\ }$}}
\newcommand{\eg}{{${ e.g.\ }$}}
\newcommand{\ea}{{${ et~al.\ }$}}
\newcommand{\hf}{{\scriptstyle{1 \over 2}}}
\newcommand{\ih}{{\scriptstyle{i \over \hbar}}}
\newcommand{\hi}{{\scriptstyle{ \hbar \over i}}}
\newcommand{\itwoh}{{\scriptstyle{i \over {2\hbar}}}}
\newcommand{\dbrst}{\delta_{BRST}}

\newcommand{\deder}[1]{{ 
 {\stackrel{\raise.1ex\hbox{$\leftarrow$}}{\delta^r}   } 
\over {   \delta {#1}}  }}
\newcommand{\dedel}[1]{{ 
 {\stackrel{\lower.3ex \hbox{$\rightarrow$}}{\delta^l}   }
 \over {   \delta {#1}}  }}
\newcommand{\dedetwo}[2]{{    { \delta {#1}} \over {   \delta {#2}}  }}
\newcommand{\dedetre}[3]{{ \left({ \delta {#1}} \over {   \delta {#2}}  
 \right)_{\! \! ({#3})} }}

\newcommand{\papar}[1]{{ 
 {\stackrel{\raise.1ex\hbox{$\leftarrow$}}{\partial^r}   } 
\over {   \partial {#1}}  }}
\newcommand{\papal}[1]{{ 
 {\stackrel{\lower.3ex \hbox{$\rightarrow$}}{\partial^l}   }
 \over {   \partial {#1}}  }}
\newcommand{\papatwo}[2]{{   { \partial {#1}} \over {   \partial {#2}}  }}
\newcommand{\papa}[1]{{  {\partial} \over {\partial {#1}}  }}
\newcommand{\papara}[1]{{ 
 {\stackrel{\raise.1ex\hbox{$\leftarrow$}}{\partial}   } 
\over {   \partial {#1}}  }}

\newcommand{\ddr}[1]{{ 
 {\stackrel{\raise.1ex\hbox{$\leftarrow$}}{\delta^r}   } 
\over {   \delta {#1}}  }}
\newcommand{\ddl}[1]{{ 
 {\stackrel{\lower.3ex \hbox{$\rightarrow$}}{\delta^l}   }
 \over {   \delta {#1}}  }}
\newcommand{\dd}[1]{{  {\delta} \over {\delta {#1}}  }}
\newcommand{\pa}{\partial}
\newcommand{\sokkel}[1]{\!  {\lower 1.5ex \hbox{${\scriptstyle {#1}}$}}}  
\newcommand{\larrow}[1]{\stackrel{\rightarrow}{#1}}
\newcommand{\rarrow}[1]{\stackrel{\leftarrow}{#1}}
\newcommand{\twobyone}[2]{\left(\begin{array}{c}{#1} \cr
                                   {#2} \end{array} \right)}
\newcommand{\twobytwo}[4]{\left[\begin{array}{ccc}{#1}&&{#2} \cr
                                  {#3} && {#4} \end{array} \right]}
\newcommand{\fourbyone}[4]{\left(\begin{array}{c}{#1} \cr{#2} \cr{#3} \cr
                                   {#4} \end{array} \right)}

\newcommand{\eq}[1]{{(\ref{#1})}}
\newcommand{\Eq}[1]{{eq.~(\ref{#1})}}
\newcommand{\Ref}[1]{{Ref.~\cite{#1}}}
\newcommand{\mb}[1]{{\mbox{${#1}$}}}
\newcommand{\equi}[1]{\stackrel{{#1}}{=}}
\newcommand{\succeqq}{\succeq}
\newcommand{\ccdot}{\cdot}
\newcommand{\qqa}{}
\newcommand{\qqb}{}
\newcommand{\qqc}{}
\newcommand{\qqd}{}
\newcommand{\yyy}{y}
\newcommand{\DD}{{\cal D}}

\newcommand{\proofbox}{\begin{flushright}
${\,\lower0.9pt\vbox{\hrule \hbox{\vrule
height 0.2 cm \hskip 0.2 cm \vrule height 0.2 cm}\hrule}\,}$
\end{flushright}}

\begin{titlepage}
\title{\Large{\bf Putting an Edge\\
to the Poisson Bracket\\ 
}}

\author{{\sc K.~Bering}
\footnote{Email address: {\tt bering@phys.ufl.edu, bering@nbi.dk}}
\\
Institute for Fundamental Theory\\Department of Physics\\
University of Florida\\Florida 32611, USA\\
{~}}

\date{MIT-CTP-2746,~~~~~ hep-th/9806249. {~~~~~}July 1998 }
\maketitle
\begin{abstract}
We consider a general formalism for treating a Hamiltonian (canonical) field 
theory with a spatial boundary. In this formalism essentially all 
functionals are differentiable from the very beginning and hence no 
improvement terms are needed. We introduce a new Poisson bracket which 
differs from the usual ``bulk'' Poisson bracket with a boundary term and 
show that the Jacobi identity is satisfied. The result is geometrized on an 
abstract world volume manifold. The method is suitable for studying systems 
with a spatial edge like the ones often considered in Chern-Simons theory 
and General Relativity. Finally, we discuss how the boundary terms may be 
related to the time ordering when quantizing. 
\end{abstract}

\bigskip
\begin{center}
Submitted to: {\it Journ.~Math.~Phys.}
\end{center}

\vspace*{\fill}

\noi
PACS number(s): 02.70.Pt, 11.10.Ef.

\end{titlepage}
\vfill
\newpage

\setcounter{equation}{0}
\section{Introduction}

\noi
Seen in the light of the renewed interest for theories where the edge plays
a prominent role, cf.\ Maldacena's Conjecture \cite{maldacena},
Carlip's and Strominger's different approaches for the microscopic counting
of states on the (inner or outer) edge of the world \cite{countingbh},  
't Hooft's and Susskind's principle of holography \cite{holo}, 
but also Chern-Simon theories \cite{witten} and General Relativity 
\cite{reggetei,brownhen} in general, there is strikingly few papers 
devoted to develope the general formalism for Hamiltonian 
(canonical) field theory in the presence of a spatial boundary. 
Here we are thinking of the fact that the usual equal-time Poisson bracket 
\beq
  \{F(t) , G(t) \}_{(0)}~\equiv~\int_{\Sigma}d^{d}x 
\dedetwo{F(t)}{\phi^{A}(x,t)} \omega^{AB}  \dedetwo{G(t)}{\phi^{B}(x,t)} 
\label{naivepb}
\eeq
fails to satisfy the Jacobi identity 
\beq
\sum_{{\rm cycl.}~ F,G,H} \{F(t) ,  \{G(t) , H(t) \} \}~=~0~,
\eeq
when space \mb{\Sigma} has a boundary \mb{\pa \Sigma \neq \emptyset},
at least if we apply the usual Euler-Lagrange formula for the
functional derivatives in \eq{naivepb}. 
(We shall show below how to ensure the differentiability of the 
functionals by using the notion of {\em higher} functional derivatives, 
so that the above violation is a fully legitimate problem to raise.)
The failure of the Jacobi identity can be seen even in the most simple
toy examples which have a non-trivial boundary. An equivalent manifestation
of this fact is that functional derivatives cease to commute when a
spatial boundary is present \cite{solovievplb}. 

\vs
\noi
The most common example of the above phenomenon is the usual 
$d$-dimensional flat space \mb{\Sigma=\R^{d}}. Here the spatial 
infinity \mb{|x|=\infty} constitute a boundary for the space.
This statement can be made precise by a so-called one-point compactification.

\vs
\noi
An often used cure is to impose conditions on the dynamical fields
\mb{\phi^{A}(x,t)} at the boundary which are consistent with the 
time-evolution. However, that approach may exclude interesting 
topological questions, such as solitonic field configurations.
Our main goal in this paper is to see how far we can get {\em without} 
imposing boundary conditions. 

\vs
\noi
On the other hand, to calm the reader who perhaps finds these facts strange, 
let us mention that if ``there is no boundary'' (for instance, think of a 
torus, or equivalently periodic boundary conditions, or even vanishing 
boundary conditions),
integrations by part inside the spatial integral does not produce boundary 
contributions, and the Jacobi identity for the above ``bulk'' Poisson 
bracket can be demonstrated after some straightforward manipulations. 

\vs
\noi
The paper is organized as follows: In next Subsection~\ref {secnewpb}, 
we present a new Poisson bracket. Thereafter, we give some further 
introductional remarks about differentiability and improvement terms.
In Section~\ref{secgenth} we give a manifest formulation of the new 
Poisson bracket, discuss the higher Euler-Lagrange derivatives
and develop a generator method (Section~\ref{generatormeth})
which in particular is suitable of handling  the arithmetic manipulations
involved in the proof of the Jacobi identity. 
The proof itself is postponed to an Appendix~\ref{comji}. 
After that we turn to the questions that naturally arise with the existence
of the new Poisson Bracket. Can it be given a geometrically covariant form
(Section~\ref{abstractmani})? How does the boundary affect the Hamiltonian 
dynamics (Section~\ref{seched})?  To answer the last question, we have
included a technical Section~\ref{secsuppth} to explain some supplementary
formalism. Finally, we discuss the role of time order in connection with 
boundary terms (Section~\ref{timeord}).

\subsection{Notation}
\label{secnota}

\noi
The \mb{\phi^{A}(x,t)}, \mb{A=1,\ldots,2N},  
denote the (bosonic) coordinate and momenta fields of the phase space. 
(Generalizations to fermionic variables are straightforward.) 
The non-degenerate symplectic structure \mb{\omega^{AB}} is for simplicity 
taken to be ultra-local and constant. This vast simplification is the
most interesting case for applications and already contain as we shall
see an interesting structure worth analyzing by itself.
Needless to say that a general field transformation \mb{\phi \to \phi'} 
will violate such assumption. A manifestly covariant formalism under
general field transformation is out of the scope of the present work.
Also we assume, to avoid technicalities, that the $d$-dimensional space 
\mb{\Sigma} can be covered by a single coordinate patch with a flat measure.
(We shall relax these assumptions in Section~\ref{abstractmani}.)
{}Furthermore, in agreement with the spirit of the Hamiltonian (canonical) 
formalism, we shall assume that the functionals \mb{F(t)} do not
contain time derivatives \mb{\left(\pa_{t}\right)^{k}\phi^{A}(x,t)} 
of the dynamical field variables \mb{\phi^{A}(x,t)}. So if one for
instance is interested in a (total) time derivative 
\mb{G(t)=\pa_{t}^{k} F(t)}, where \mb{F(t)} 
contains no time derivatives, one should study \mb{F(t)} instead of 
\mb{G(t)}, etc. Finally, we assume that there is no temporal boundaries.
This being said, time $t$ plays no active role, and we can suppress the 
time variable $t$ in what follows.

\subsection{New Poisson Bracket}
\label{secnewpb}

\noi
As mentioned in the introduction the bulk Poisson bracket \eq{naivepb} does 
not satisfy the Jacobi identity. (A purist would perhaps then claim that 
\eq{naivepb} does not qualify for being called a Poisson bracket at all! 
However,  we shall continue to call it a Poisson bracket.)
Knowing that the failure of the Poisson bracket \eq{naivepb} is at most a 
total derivative term, it is natural to speculate 
whether one can add a boundary contribution \mb{B(F,G)} to this bulk Poisson 
bracket, 
\beq
 \{F , G \}~=~\{F , G \}_{(0)} + B(F,G) ~,
\label{fullpb}
\eeq
so that the Jacobi identity is satisfied even in the presence of a boundary. 
In fact, this is so. We find that the following boundary term
\beq
B(F,G)~=~ \sum_{k\neq 0} \int_{\Sigma} d^{d}x ~\pa^{k} \left[  
\dedetwo{F}{\phi^{A(k)}(x)} \omega^{AB} \dedetwo{G}{\phi^{B}(x)} \right]
~-~ (F \leftrightarrow G)
\label{bboundterm}
\eeq
does the job.
We have employed a multi-index notation: For instance 
the index 
\beq
k~=~(k_{1}, \ldots, k_{d}) \in \N_{0}^{d}\backslash \{(0, \ldots,0) \} ~,
~~~~~~~\N_{0}~\equiv~\{0,1,2,\ldots\}~,
\label{kayindex}
\eeq
runs over the $d$-dimensional non-negative integers (except the origo), and
\beq
   \pa^{k}~=~\pa_{1}^{k_{1}} \cdots \pa_{d}^{k_{d}}~,~~~~~~~~~~~~
\pa_{i}\equiv \papa{x^{i}}~.
\eeq
(The main features of the construction are already present in the
dimension \mb{d=1} case. A first-time reader will not miss the essential 
points by treating the multi-index \mb{k} as an integer, \ie letting 
\mb{d=1}.) More importantly, in a perhaps conceptionally dangerous -- but in
practice convenient -- notation, the
\beq
\dedetwo{F}{\phi^{A(k)}(x)}~,~~~~~~k \in \N_{0}^{d}~,
\eeq
denote the {\em higher} functional derivatives of $F$ of order $k$. Here 
\beq
   \dedetwo{F}{\phi^{A(k=0)}(x)}~\equiv~ \dedetwo{F}{\phi^{A}(x)}
\label{usualfuncderiv}
\eeq
is the usual functional derivative.
In general, the higher functional derivatives are required to satisfy
\beq
\delta F ~=~ \int_{\Sigma} d^{d}x ~ \sum_{k=0}^{\infty} \pa^{k} \left[ 
\dedetwo{F}{\phi^{A(k)}(x)} \delta \phi^{A}(x) \right]
\label{highfuncderiv}
\eeq
for arbitrary infinitesimal variations 
\mb{\phi^{A}(x) \to \phi^{A}(x)+ \delta \phi^{A}(x)}. In particular,
the variations \mb{\delta \phi^{A}(x)} are not restricted at the 
boundary.

\vs
\noi
A quick estimate shows that if the entries contain spatial derivatives 
\mb{\pa_{i}\phi^{A}(x)} of the dynamical field variables \mb{\phi^{A}(x)} to 
order $N$, then the full Poisson Bracket \eq{fullpb} contains spatial 
derivatives up to order $3N$. Hence the algebra \mb{{\cal A}_{N=0}} of 
functionals with no spatial derivatives closes on the new Poisson bracket. 
However, physical interesting theories usually have functionals with up to 
\mb{N=1} spatial derivative, \ie  they belong to the class 
\mb{{\cal A}_{N=1}}. This class \mb{{\cal A}_{N=1}} of functionals does not 
close on the new nor on the bulk Poisson bracket, and this is the main 
reason why we are force into summing over the index lattice \eq{kayindex}.

\vs
\noi
The idea of adding surface contribution is far from new.
In a seminal work Regge and Teitelboim \cite{reggetei} emphasized the
importancy of having a boundary term in the action of canonical
general relativity. However, they did strangely enough not add surface 
contributions to the Poisson bracket. Lewis, Marsden, Mongomery and 
Ratiu \cite{lmmr} considered a truncated version 
of \eq{bboundterm} where only the terms with \mb{|k|=1} are present.
Because they didn't add the \mb{|k|>1} terms, they needed to impose
additional boundary conditions to ensure the Jacobi identity.
The first successful attempt to remedy this was conducted in 1993 by Soloviev 
\cite{soloviev}. In our notation, his solution \cite[formula(3.4)]{soloviev} 
reads 
\beq
 \{F , G \}~=~
\sum_{k,\ell=0}^{\infty} \int_{\Sigma} d^{d}x ~\pa^{k+\ell} \left[  
\dedetwo{F}{\phi^{A(k)}(x)} \omega^{AB} 
\dedetwo{G}{\phi^{B(\ell)}(x)} \right]~.
\label{solovievbracket}
\eeq 
It is easy see that his bracket is different from our solution 
\eq{bboundterm}.
It would be interesting to know whether his bracket supports a
manifest formulation (see Subsection~\ref{secmannisform} below), or
more generally, if it is independent of the representative for the 
higher functional derivatives. After the first preprint of this
paper appeared, Soloviev has made a comparison \cite{solovievcomp} 
of the two solutions \eq{bboundterm} and \eq{solovievbracket}. 
We shall in this paper concentrate fully on the solution \eq{bboundterm}.

\subsection{Review of Differentiability and Improvement Terms}

\noi
The classical point of view \cite{reggetei} on the problem with the Jacobi 
identity has been to view this as not so much a problem of the Poisson 
bracket itself, but rather that functional derivatives in general in the 
case of a non-trivial boundary \mb{\pa \Sigma \neq \emptyset} are 
ill-defined when the functional, say $F$, depends on the spatial derivatives 
\mb{\pa_{i}\phi^{A}(x)} of the dynamical field variables 
\mb{\phi^{A}(x)}. In this case there does not always exist functions
\mb{f_{A}(x)}, such that the change in the functional $F$ is fully described 
by
\beq
 \delta F ~=~ \int_{\Sigma} d^{d}x ~f_{A}(x)~ \delta \phi^{A}(x)~,
\label{traddiff}
\eeq  
for an arbitrary infinitesimal variation \mb{\delta\phi^{A}(x)}. 
(Of course in the affirmative case, we usually call \mb{f_{A}(x)} 
the functional derivatives of $F$.)

\vs
\noi
{\bf Example.} {\em Consider an interval \mb{\Sigma=[a,b]} and
the toy functional 
\beq
 F~=~\int_{a}^{b} dx~ \pa^{2} \phi(x)~=~\pa \phi(x)|^{x=b}_{x=a}~.
\eeq
The variation \mb{\delta F} can be identically rewritten as
\beq
\delta F ~=~ \int_{\Sigma} 
\left(\delta_{\Sigma}(x,b)-\delta_{\Sigma}(x,a)\right)  \pa \delta \phi(x)
\eeq
To bring \mb{\delta F} of the form \eq{traddiff}, one is tempted to do an 
integration by part. But this does not help us, partly because of the total 
derivative term does not vanish on the boundary. (Remember that we do not 
want to impose boundary conditions on the field \mb{\phi}, cf.\ the 
Introduction.) In fact, it is not apriori clear what should be meant (viewed 
separately) by any of the \mb{2 \times 2} terms  arising from such an 
integration by part. So the traditional functional derivative is ill-defined.
These difficulties should be compared with the ease that the same variation 
\mb{\delta F} is described by the higher functional derivatives 
\eq{highfuncderiv}
\beq
 \dedetwo{F}{\phi^{(k)}(x)}~=~\delta_{k}^{2} ~.
\eeq  
} 

\vs
\noi
This is the general picture: A traditional functional derivative is often
ill-defined or have a very singular behavior at the boundary. (And at this 
point we haven't even touched the problems of building up the Poisson bracket 
\eq{naivepb} out of two highly singular functional derivatives, \ie 
multiplying two delta distributions together, both of which typically have 
support on the boundary. We should also mention that authors 
for such reasons often additionally require the functional derivatives 
\mb{f_{A}} in Eq.\ \eq{traddiff} to be continuous functions.) 
In any case, this makes the traditional definition \eq{traddiff} 
not very suitable for systems with a boundary.

\vs
\noi
Let us mention an important algebra \mb{{\cal A}_{0}} of functionals, that 
{\em are} differentiable in this traditional sense \eq{traddiff}, namely 
those functionals that do not depend on the spatial 
derivatives \mb{\pa_{i}\phi^{A}(x)} of the dynamical field variables 
\mb{\phi^{A}(x)}. They form an algebra under the Poisson bracket. (In this 
algebra \mb{{\cal A}_{0}} the bulk Poisson bracket \eq{naivepb} and the 
full Poisson bracket \eq{fullpb} coincide.)

\vs
\noi
The traditional 
cure in case of an ill-defined derivative, is to improve the functional 
\mb{F\leadsto F_{\rm impr}} with a boundary term, a so-called improvement 
term, so that the derivative becomes smooth and well-defined. The drawback 
is of course that we are then studying a different functional than we 
originally started out with! Typically, one meets the following 
preparation of an observable in the literature:
A function \mb{f(x)}\mb{=}\mb{f(\pa^{k} \phi(x),x)} is smeared with
a ``test function'' \mb{\eta(x)} into a functional of the type
\mb{F[\eta]=\int_{\Sigma} d^{d}x~f(x)~ \eta(x)}. It is then improved 
\mb{F[\eta] \leadsto F_{\rm impr}[\eta] \in {\cal A}_{0}} so that it belongs
to the above mentioned class \mb{{\cal A}_{0}} by recasting 
all the spatial derivatives to hit the test function. 

\vs
\noi
We will bypass all this, \ie the  bottle-neck of \eq{traddiff},
by using more functions (the higher functional derivatives) in the 
description of an arbitrary variation \mb{\delta F}. The
format of \eq{highfuncderiv} is so broad, that it essentially covers
all interesting functionals, which do not contain time-derivatives,
cf.\ the discussion in Subsection~\ref{secnota}. (One could of course 
give meaning to differentiation of a functional with temporal derivatives 
simply by brute force extending the multi-index $k$ in \eq{highfuncderiv} 
from $d$ dimensions to \mb{d+1} dimensions. Although relevant for so-called
covariant formulations (covariant in the sense that time and space are 
treated on equal footing), this is not in the line of the Hamiltonian 
theories, and hence not something we will pursue in this paper.)

\setcounter{equation}{0}
\section{General Formalism}
\label{secgenth}

\subsection{Partial Derivatives of a Functional}

\noi
Let us describe the higher partial derivatives of a functional $F$.
(They are not to be confused with the usual higher partial derivatives of a 
function, although they are related.) In fact, they are objects, given the 
suggestive notation 
\beq
    \papatwo{F}{\phi^{A(k)}(x)}~, 
\label{papanice}
\eeq
that satisfies 
\beq
\delta F~=~\int_{\Sigma} d^{d}x ~\sum_{k=0}^{\infty} 
\papatwo{F}{\phi^{A(k)}(x)}~\pa^{k} \delta \phi^{A}(x)~,
\label{highpartderiv}
\eeq
for arbitrary variations \mb{\delta \phi^{A}(x)}.
If this notation \eq{papanice} in the future leads to ambiguities, 
we will specify whether we mean partial differentiation wrt.\ a function 
or a functional. Usually the context will exclude one of the possibilities.
In fact, in this article we will often use the notation
\beq
  P_{A(k)}f(x)~\equiv~\papatwo{f(x)}{\phi^{A(k)}(x)}~.
\eeq
for the usual (higher) partial derivative for a function 
\mb{f(x)= f(\pa^{k}\phi(x),x)}.

\subsection{Manifest Formulation}
\label{secmannisform}

\noi
The set of the higher functional (and the partial) derivatives may not
be uniquely defined, so one may worry that the full Poisson bracket
given by \eq{fullpb} and \eq{bboundterm} depends on the choice of the 
representatives for the higher functional derivatives. The answer is
that it is independent. This follows from the manifest formulation given below.

\vs
\noi
We begin by giving a more useful definition of the (usual) functional 
derivatives \eq{usualfuncderiv} than the traditional definition, 
cf.\ Eq.\ \eq{traddiff}. The 
differential \mb{\delta F=\delta F[\phi, \delta\phi]} of \mb{F} is 
assumed to split
\beq
  \delta F~=~d F + \pa F
\label{split}
\eeq
into a bulk integral 
\beq
 d F~=~d F[\phi, \delta\phi]~=~\int_{\Sigma} d^{d}x ~\dedetwo{F}{\phi^{A}(x)}~
 \delta \phi^{A}(x)
\eeq
 and 
a boundary integral 
\beq
 \pa F~=~\pa F[\phi, \delta\phi]~=~ \sum_{i=1}^{d}\int_{\Sigma} d^{d}x~\pa_{i}
~\hat{F}^{i}_{A} \delta \phi^{A}(x)~, \label{boundint}
\eeq
where \mb{\hat{F}^{i}_{A}} in general can be differential operators acting
on \mb{\delta \phi^{A}}. That the 
integral \eq{boundint} is a boundary integral is justified by the divergence 
theorem. If we furthermore require the (usual) functional derivative to be 
a contineous function, this function together with the 
above spilt \eq{split} are uniquely defined. (We stress that the uniqueness 
of the (usual) functional derivative is jeopardized if for instance it 
contains a delta distribution with support on the boundary. This is ruled 
out by requiring continuity. See also the uniqueness discussion in the next
Section.)

\vs
\noi
The bulk Poisson bracket \mb{\{\cdot, \cdot \}_{(0)}} is now well-defined 
by the Eq.\ \eq{naivepb}, and the full Poisson bracket Eq.\ \eq{fullpb} 
differs from this by a boundary term
\beq
  B(F,G) ~=~ \pa F\left[ \phi,\delta\phi\!=\! \{\phi, G \}_{(0)} \right] 
~-~ (F \leftrightarrow G)~.
\eeq
At this point the reader can merely take
\mb{\delta\phi\!=\! \{\phi, G \}_{(0)}} as being a convenient short hand 
notation for
\beq
\delta\phi^{A}(x)~=~ \omega^{AB}\dedetwo{G}{\phi^{B}(x)} ~. 
\eeq
There is two obvious ansatzes for the differential operator 
\mb{\hat{F}^{i}_{A}}. Either with the spatial derivatives ordered to the 
right of some coefficient functions (also called the normal order of
\mb{x} and \mb{\pa}). Or vice-versa. In the former case
\beq
   \hat{F}^{i}_{A}~=~\sum_{k=0}^{\infty} 
\papatwo{{}^{i}F}{\phi^{A(k)}}~\pa^{k} ~,
\label{highpartvectdef}
\eeq
we call the coefficient functions for the higher partial vector derivatives.
The name ``vector'' refers to the \mb{i}-index. In the latter case
\beq
   \hat{F}^{i}_{A}~=~\sum_{k=0}^{\infty} 
\pa^{k} \dedetwo{{}^{i}F}{\phi^{A(k)}}~.
\label{highfuncvectdef}
\eeq
the coefficient functions are called the higher functional vector derivatives.

\subsection{Uniqueness of the Higher Derivatives}

\noi
Up to now, we have only characterized the higher functional (partial, 
vector) derivatives of a functional \mb{F} in a descriptive manner.
The question of existence yields rather mild conditions, that we shall not 
be concerned about. The issue of uniqueness is a much more interesting 
question. The ambiguity in the choice is most clearly displayed via the
higher partial vector derivatives:

\vs
\noi
{\bf Uniqueness Theorem}. {\em Assume that there is given a sequence of 
contineous functions \mb{f_{A}} and \mb{f^{i}_{(k)A}} that all vanish 
identically with the exception of a finite number and such that for an
arbitrary variation \mb{\delta \phi}
\beq 
  0 ~\equiv~ \int_{\Sigma} d^{d}x ~f_{A}(x)~ \delta \phi^{A}(x) 
+\sum_{i=1}^{d}\int_{\Sigma} d^{d}x~\pa_{i}  
\sum_{k=0}^{\infty} f_{A(k)}^{i}(x)~\pa^{k} \delta \phi^{A}(x)~.
\eeq
Then \mb{f_{A}\equiv 0} in the entire space \mb{\Sigma} and the (higher) 
partial vector derivatives \mb{f_{A(k)}^{i}} are tangential to the boundary 
\mb{\pa\Sigma}. In detail,
\beq 
\forall x \in \pa \Sigma~:~ 
\vec{f}_{A(k)}(x)\equiv \left(f_{A(k)}^{1}(x), \ldots,f_{A(k)}^{d}(x)\right)  
\in T_{x}(\pa \Sigma)~.
\eeq }
That the first function \mb{f_{A}\equiv 0} vanishes is just a restatement 
of the uniqueness of the (usual) functional derivative, while the boundary
condition reflects that the higher partial vector derivatives may be 
modified with a vector field that doesn't locally carry a boundary flux.

\subsection{Relations among the Different Kinds of Higher Derivatives}

\noi
In order to get an idea of how ambiguous the other variational discriptions  
are, let us give some maps between the mentioned choices of the higher 
functional (partial, vector) derivatives. We start with a bijective 
correspondance between the two scalar definitions \eq{highfuncderiv} and 
\eq{highpartderiv} of the higher derivatives. If there is given a sequence 
of the higher partial derivatives satisfying the definition 
\eq{highpartderiv}, then
\beq
   \dedetwo{F}{\phi^{A(k)}(x)}
~=~ \sum_{m \geq k} \twobyone{m}{k} 
(-\pa)^{m-k} \papatwo{F}{\phi^{A(m)}(x)}~,~~~~~~
\twobyone{m}{k}~=~\twobyone{m_{1}}{k_{1}} \cdots \twobyone{m_{d}}{k_{d}}~.   
\label{higherhigherfunc}
\eeq
satisfies the definition \eq{highfuncderiv}.
On the other hand, starting from the point of view of the higher functional 
derivatives, we get a solution to the higher partial derivatives by
\beq
   \papatwo{F}{\phi^{A(k)}(x)}
~=~ \sum_{m \geq k} \twobyone{m}{k} 
\pa^{m-k} \dedetwo{F}{\phi^{A(m)}(x)}~.  
\label{higherhigherpart} 
\eeq
The variational descriptions \eq{highfuncderiv} and \eq{highpartderiv} 
coincides because of the $x$-pointwise identity 
\beq
\sum_{k=0}^{\infty} \papatwo{F}{\phi^{A(k)}(x)}~
\pa^{k} \delta \phi^{A}(x)   ~=~  
\sum_{k=0}^{\infty} \pa^{k} \left[ \dedetwo{F}{\phi^{A(k)}(x)} 
\delta \phi^{A}(x) \right]~.
\label{arewerollingsomemore}
\eeq
{}For a proof, see the equation \eq{wearerolling} in the Appendix.
Similarly, one may transform bijectively between the two vectorial 
definitions by use of the corresponding relations
\beq
   \dedetwo{{}^{i}F}{\phi^{A(k)}(x)}
~=~ \sum_{m \geq k} \twobyone{m}{k} 
(-\pa)^{m-k} \papatwo{{}^{i}F}{\phi^{A(m)}(x)}~,~~~~~~~    
 \papatwo{{}^{i}F}{\phi^{A(k)}(x)}
~=~ \sum_{m \geq k} \twobyone{m}{k} 
\pa^{m-k} \dedetwo{{}^{i}F}{\phi^{A(m)}(x)}~.   
\label{higherhighervect}
\eeq
Not surprisingly, as the vectorial definitions \eq{highpartvectdef} and
\eq{highfuncvectdef} carry the most indices, they have the greatest 
flexibility in representing a solutions.
We may convert from the higher vector to the higher scalar derivatives via 
the formulas (for \mb{k \neq 0}):
\beq
 \papatwo{F}{\phi^{A(k)}(x)}~=~
\sum_{i \in I(k)} \papatwo{{}^{i}F}{\phi^{A(k-e_{i})}(x)}
+\sum_{i =1}^{d}\pa_{i}\papatwo{{}^{i}F}{\phi^{A(k)}(x)}~,~~~~~~~~~
 \dedetwo{F}{\phi^{A(k)}(x)} ~=~ 
\sum_{i \in I(k)} \dedetwo{{}^{i}F}{\phi^{A(k-e_{i})}(x)}~.
\eeq
Here \mb{e_{i}\equiv(0,\ldots,0,1,0,\ldots,0)} is the $i$'th unit vector
in the index lattice and 
\mb{I(k)\equiv\left\{  \left. i=1,\ldots,d~\right|~ k_{i} \neq 0\right\} }.
Going from the higher scalar to the higher vector derivatives is not
unique. A natural choice is for the functional derivatives are 
\beq
 \dedetwo{{}^{i}F}{\phi^{A(k)}(x)}~=~\frac{1}{|I(k+e_{i})|}
\dedetwo{F}{\phi^{A(k+e_{i})}(x)}~.
\eeq
We leave out the corresponding relation between the partial derivatives
to carry on with our main application: Local field theories.

\subsection{Local Field Theory}
\label{seclft}

\noi
Let us restrict ourselves to local field theories, \ie
all functionals can be expressed as an integral
\beq
   F~=~\int_{\Sigma} d^{d}x ~f(x)~,~~~~~~~~~~~~
f(x)~\equiv ~f\left(\pa^{k}\phi(x),x \right)~, 
\label{locfunc}
\eeq  
where $k$ runs over a {\em finite} subset of \mb{\N_{0}^{d}}.
Note that we have allowed for explicit $x$-dependence in $f$.
It essentially costs no extra work, and it becomes important later on.
We shall postpone the analysis of functionals that depends on external 
space-points to a later section (Section~\ref{secsuppth}), partly 
because it would be notational inconvenient to address those now. 

\vs
\noi
As mentioned before the (higher) partial derivatives of $F$ need not be
unique. Our strategy will be to ``lower'' the definition from the level of
integrals to the level of integral {\em kernels}. In other words, this 
means that if a functional has different integral kernel representations, 
this may lead to different definitions of the higher derivatives.  
For instance, at the case at hand, \ie of a {\em local} functional 
\eq{locfunc} with a given integral kernel \mb{f}, there is one natural 
candidate
\beq
\papatwo{F}{\phi^{A(k)}(x)}~=~ P_{A(k)}f(x)~,
\label{highpa}
\eeq
that fits the relation \eq{highpartderiv}. So strictly speaking, the 
(higher) partial derivatives are really  (higher) partial derivatives
of the kernels, although we will not indicate this explicitly in
the notation. In the same fashion the distinguished candidate for the 
functional derivatives becomes the higher Euler-Lagrange derivatives:
\beq
\dedetwo{F}{\phi^{A(k)}(x)}~=~ 
E_{A(k)}f(x)~\equiv~ \sum_{m \geq k} \twobyone{m}{k} 
(-\pa)^{m-k} P_{A(m)}f(x)~.
\label{highel}
\eeq
For a mathematical textbook on higher Euler-Lagrange derivative, see for 
instance Olver \cite[p.365-367]{olver}. Note that the \mb{m}-summation in 
\eq{highel} terminates after finite many terms in case of a local 
functional, so that we do not have to worry about convergence of the sum.
Let us simply use \eq{highpa} (and \eq{highel}) as the working definitions 
for the local functionals. After all, our primary goal is to prove the
Jacobi identity for the full Poisson Bracket, and this does not depend on 
the choice of the representatives for the higher derivatives. It is evident 
from the natural solution \eq{highpa} (and \eq{highel}) of the (higher) 
derivatives, that in practice all the local functionals that one encounters 
in physics are differentiable.

\subsection{Ultra-local Poisson Bracket} 

\noi
Having restricted ourselves to the ultra-local case, let us
for each pair of local functional 
\beq
F~=~\int_{\Sigma} d^{d}x ~f(x)~~~~~ {\rm and}~~~~~ 
G~=~\int_{\Sigma} d^{d}x ~g(x) 
\eeq
define a $x$-pointwise version
\beq
 \{f, g\}(x) ~\equiv~ \{f, g\}_{(0)}(x)+ B(f,g)(x)
\eeq
 of the Poisson bracket 
\beq
\{F,G\}~=~ \int_{\Sigma} d^{d}x ~\{f, g \}(x)~=~ \{F, G\}_{(0)}+ B(F,G)~. 
\eeq
Namely define
\bea
 \{f, g\}_{(0)}(x)&\equiv&E_{A(0)}f(x) ~\omega^{AB}~  E_{B(0)}g(x) \cr
 B(f,g)(x)&\equiv&\sum_{k\neq 0} ~\pa^{k} \left[ E_{A(k)}f(x) 
~\omega^{AB}~  E_{B(0)}g(x) \right] ~-~ (f \leftrightarrow g)~.
\eea
This means that the full $x$-pointwise Poisson bracket reads
\bea
  \{f, g\}(x) &=& 
\sum_{k=0}^{\infty} ~\pa^{k} \left[ E_{A(k)}f(x) ~\omega^{AB}~  
E_{B(0)}g(x) \right] - \frac{1}{2} \{f, g\}_{(0)}(x)
~-~ (f \leftrightarrow g) \cr
 &=&\sum_{k=0}^{\infty} ~ P_{A(k)}f(x) ~\omega^{AB}~  
\pa^{k} E_{B(0)}g(x) -  \frac{1}{2} \{f, g\}_{(0)}(x)
~-~ (f \leftrightarrow g)~.
\label{fullpbx}
\eea
The last equality in \eq{fullpbx} follows from equation 
\eq{arewerollingsomemore}. We can now conduct our analysis $x$-pointwisely. 
In the next Section~\ref{generatormeth} we shall suppress the space point 
\mb{x \in \Sigma}.

\setcounter{equation}{0}
\section{Generator Methods}
\label{generatormeth}

\subsection{Heisenberg Algebra}

\noi
Due to the quite heavy combinatorics involved in the proof of the Jacobi 
identity, it is useful to map the above problem into a simpler and in fact 
well-studied object, namely the Heisenberg algebra. Although the actual
proof is presented in the Appendix, we find the central idea, while
perhaps not entirely original, is quite important, so we will present it here.
{}For a recent exposition of Fock space methods for variational systems, 
see also \cite{grigore}.

\vs
\noi
Let us study the interplay between the partial and the spatial derivatives. 
The higher partial derivatives \mb{P_{A(k)}} commute among each other, 
but they do not commute with the (total) spatial derivatives 
\beq
\pa_{i}~=~ \phi^{A(k+e_{i})} P_{A(k)} + \pa_{i}^{{\rm explicit}}~,
\label{totalspatialder}
\eeq
where \mb{e_{i}\equiv(0,\ldots,0,1,0,\ldots,0)} is the $i$'th unit vector
in the index lattice. More precisely, we have
\beq
  P_{A(k)}~\pa^{n}~=~
\sum_{m=0}^{{\rm min}(k,n)}\twobyone{n}{m} \pa^{n-m}P_{A(k-m)}~.
\label{combdisentang}
\eeq
The main idea is to simulate this complicated disentanglement
formula with the help of a Heisenberg algebra. Let us introduce 
abstract (bosonic) algebra elements \mb{Y^{i}_{A}} that obey 
the following Heisenberg algebra commutator relations
\beq
  [Y^{i}_{A},\pa_{j}]~=~\delta^{i}_{j}~,~~~~~~~~ 
[Y^{i}_{A},Y^{j}_{B}]~=~0~,~~~~~~~~
[\pa_{i},\pa_{j}]~=~0~.
\label{heisalg}
\eeq
The third equation is not a definition, but is a well-known consequence 
of \eq{totalspatialder}.
It is a remarkable fact that we can mimick the non-commuting behavior
of formula \eq{combdisentang} by formally writing the higher partial 
derivatives as a product
\beq
  P_{A(k)}~\equiv~P_{A}\frac{Y_{A}^k}{k!}
\label{formalassign}
\eeq
of the \mb{Y^{i}_{A}} algebra elements and what is basically reduced to 
be a passive spectator in what follows, namely
\beq
 P_{A}~\equiv~\papa{\phi^{A}}~.
\eeq
We take \mb{P_{A}} to commute with everything:
\beq
  [P_{A},Y^{i}_{B}]~=~0~,~~~~~~~~  [P_{A},\pa_{i}]~=~0~,~~~~~~~~  
[P_{A},P_{B}]~=~0~.
\eeq
Above we have adapted the following multi-index conventions:
\beq
Y_{A}^{k}~=~ (Y^{1}_{A})^{k_{1}} \cdots (Y^{d}_{A})^{k_{d}}~,~~~~~~~~
k!~=~k_{1}! \cdots k_{d}!~.
\eeq
That the Heisenberg algebra \eq{heisalg}
with the formal assignment \eq{formalassign} really reproduces 
\eq{combdisentang} is proven in the Appendix~\ref{varid}, 
see equation \eq{reallyheis}. 
The proof becomes very simple once we adapt the 
generator techniques of the next section.

\subsection{Generator Methods}

\noi
As a second computational improvement, it is useful to hide the integer 
indices inside generating functions which depend on continuous 
parameters \mb{q_{i}}, \ie we shall sum up in generalized Fourier series.
We implement this program for the (higher) partial and the (higher) 
Euler-Lagrange derivatives, respectively, as follows
\bea
  P_{A}(q)&\equiv&P_{A(k)}q^{k}~=~P_{A}~e^{qY_{A}}~, \cr
E_{A}(q)&\equiv&E_{A(k)}q^{k}~=~~:\exp\left[(q-\pa)Y_{A}\right]P_{A}:
~~=~\exp\left[-\pa \papa{q}\right]P_{A}(q)~.
\label{peq}
\eea
{}From now on we will implicitly imply summation over repeated multi-indices
\mb{k \in \N_{0}^{d}}. In fact, we may view the multi-indices sums as 
running over the entire $d$-dimensional integer lattice \mb{k \in \Z^{d}}
by simply declaring that objects like
\beq
     Y^k_{A}~,~~\pa^{k}~,~~P_{A(k)}~,~~ E_{A(k)}~,~~
\eeq
are zero if $k$ is outside the original non-negative $d$-dimensional 
quadrant \mb{\N_{0}^{d}}. The next-to-last equality in \eq{peq} follows 
from the mere definition of the (higher) Euler-Lagrange derivatives 
\eq{highel}, once we have declared the following normal ordering 
prescription:
\beq
 :Y^{i}_{A}\pa_{j}:~~=~~:\pa_{j}Y^{i}_{A}:~~=~~\pa_{j}Y^{i}_{A}~.
\eeq

\subsection{Fourier Transform}
\label{secfouriertrans}

\noi 
As a third computational improvement, let us Fourier transform
the variables \mb{q_{i}} to variables \mb{y^{i}}. 
\bea
  P_{A}(y)&\equiv&\int d^{d}q~e^{-qy} P_{A}(q)~=~P_{A}~\delta(Y_{A}-y)~, \cr
E_{A}(y)&\equiv&\int d^{d}q~e^{-qy} E_{A}(q) ~=~e^{-\pa y} P_{A}(y)
~=~ P_{A}~\delta(Y_{A})~e^{-\pa y}~.
\label{pey}
\eea
With the theory developed so far, we have reached the second objective
of this paper (the first objective being to give the form of the full 
Poisson bracket
\eq{fullpb}), namely achieved a formalism that are capable of giving a 
short proof of the Jacobi identity. For the proof itself, see the 
Appendix~\ref{comji}. Note that because the Poisson bracket is independent
of the actual choice of representatives for the higher functional (partial) 
derivatives, it implies no limitation that we use the natural choice 
\eq{highel} and \eq{highpa}. We will now turn to the question of geometrizing 
the Poisson bracket.



\setcounter{equation}{0}
\section{Abstract Manifolds}
\label{abstractmani}

In this section we formulate the results obtained so far in a geometrically 
covariant manner independent of the choice of coordinates. More precisely, 
the construction is generalized from a $d$-dimensional subset 
\mb{\Sigma \subseteq \R^{d}} (where the space and the chart are identified) 
to an abstract $d$-dimensional manifold \mb{\Sigma} with spatial covariant 
derivatives \mb{D_{i}} and a $d$-dimensional volume density \mb{\rho}. 
In other words, the spatial derivatives \mb{\pa_{i}} are replaced by 
covariant derivatives (let us indicate this in an oversimplified way as 
\mb{D_{i}=\pa_{i}+\Gamma_{i}}), and the trivial measure \mb{d^{d}x} is  
replaced by \mb{\rho d^{d}x}. We will assume that \mb{D_{i}=D_{i}(x)} and 
\mb{\rho=\rho(x)} do not depend on the dynamical fields 
\mb{\phi^{A}(x,t)} nor on time $t$. We do {\em not} assume that the
volume density is covariantly preserved, \ie that 
\mb{D_{i}\rho\equiv(\pa_{i}-\Gamma_{ik}^{k})\rho=0}.

\vs
\noi
In passing from derivatives
\mb{\pa_{i}} to covariant derivatives \mb{D_{i}},
we face the main complication compared to the flat case.
In general, the spatial covariant derivatives do not commute,
when the curvature is non-vanishing. We have
\beq
   [D_{i}, D_{j}] ~=~ [D_{i}^{{\rm explicit}}, D_{j}^{{\rm explicit}}]~,
\eeq
where the (total) covariant  derivative \mb{D_{i}} is given as
\beq
D_{i}~=~ \phi^{A(i+K)} P_{A(K)} + D_{i}^{{\rm explicit}}~.
\label{totalcovspatialder}
\eeq
The index structure of the first term in \eq{totalcovspatialder} will
be explained below. In the general case of non-vanishing curvature, 
one can proceed by declaring that the functionals of the theory depend on 
{\em ordered} tuples of covariant derivatives 
\mb{\phi^{A(K)}(x)=D^{K}\phi^{A}(x)} 
of the dynamical fields \mb{\phi^{A}(x)}, rather than only {\em unordered} 
sets of derivatives \mb{\phi^{A(k)}(x)}. An ordered tuple \mb{K} is of 
the form
\beq
K~=~(k_{1},\ldots, k_{|K|}) ~\in~ \{1, \ldots, d \} \times \ldots \times 
  \{1, \ldots, d \} ~,
\eeq
where \mb{d} is the space dimension, \ie the dimension of \mb{\Sigma}. 
We have given a resum\'{e} of the calculus of ordered tuples in
the Appendix~\ref{appcalcorderedtuples}.
All formulas carries over to the curved case in  essentially the same 
format. However, there are some notable differences that we now stress.
The description of the higher derivatives  \eq{highpartderiv}, 
\eq{highfuncderiv}, \eq{higherhigherfunc} and \eq{higherhigherpart}
are replaced with 
\bea
\delta F&=&\int_{\Sigma} \rho(x)  d^{d}x \sum_{K=\emptyset}^{\infty} 
\papatwo{F}{\phi^{A(K)}(x)}~D^{K} \delta \phi^{A}(x) \cr
 &=&\int_{\Sigma}\rho(x)   d^{d}x \sum_{K=\emptyset}^{\infty} D^{K} \left[ 
\dedetwo{F}{\phi^{A(K)}(x)}~\delta \phi^{A}(x) \right]~, \cr
 \dedetwo{F}{\phi^{A(K)}(x)}
&\equiv& \sum_{M \succeq K} ~ 
(-D)^{M^{t}\div K^{t}} \papatwo{F}{\phi^{A(M)}(x)}~, \cr
\papatwo{F}{\phi^{A(K)}(x)}
&=& \sum_{M \succeq K} ~ 
D^{M\div K} \dedetwo{F}{\phi^{A(M)}(x)}~.
\label{ordhighderiv}
\eea
For details concerning the notation in \eq{ordhighderiv}, 
see definitions in Appendix~\ref{secwordss}.  
The (higher) functional and partial derivatives inherits tensor properties, 
if $F$ is covariant. So the formulas are covariant.
The bulk and the boundary term of the Poisson bracket reads
\bea
\{ F, G \}_{(0)} 
&=& \int_{\Sigma} \rho(x) d^{d}x 
\dedetwo{F}{\phi^{A}(x)} \omega^{AB}  \dedetwo{G}{\phi^{B}(x)}  \cr
B(F,G)&=& \sum_{K\neq \emptyset} 
\int_{\Sigma}\rho(x)   d^{d}x ~D^{K} \left[  
\dedetwo{F}{\phi^{A(K)}(x)} \omega^{AB} \dedetwo{G}{\phi^{B}(x)} \right]
~-~ (F \leftrightarrow G)~.
\label{fullpb2}
\eea
Note the apparent asymmetry between the two last formulas in 
\eq{ordhighderiv}  with a transposition $t$ of the order of covariant 
derivatives in the third equation. As a rule of thumb one may say that 
the spatial derivatives \mb{(-\pa)^{k}} in the flat metric formulation 
becomes \mb{(-D)^{K^{t}}} in the covariant formulation. This generic 
feature carries over to the generator formalism \eq{peq}:
\bea
  P_{A}(q)&\equiv&P_{A(K)}q^{K}~=~P_{A}~e^{q\ccdot Y_{A}}
~~=~\exp\left[ D\ccdot \papa{q}\right] E_{A}(q)~, \cr
E_{A}(q)&\equiv&E_{A(K)}q^{K}
~~=~\exp\left[(-D)^{t}\ccdot \papa{q}\right] P_{A}(q)~.
\label{curvedpeq}
\eea
Let us note that the \mb{q_{i}}'s (besides commuting with everything else) 
does {\em not} commute among themselves. More precisely, they are freely 
generated. 
This is necessary in order not to loose information about the operator 
ordering when passing to the generating functions. Also the Fourier 
transform \mb{q^{i} \to y_{i}}, cf.\ Section~\ref{secfouriertrans}, 
can be given sense in the non-commutative case.

\vs
\noi
The replacement of the disentanglement formula \eq{combdisentang} becomes
\beq
  P_{A(K)}~D^{N}~=~
\sum_{M=\emptyset}^{M \leq K, M \preceq N} D^{N\div M} P_{A(-M+K)}~.
\label{covcombdisentang}
\eeq
We are able to define contravariant 
elements \mb{Y_{A(K)}}, such that
\beq
  P_{A(K)}~\equiv~P_{A}~\frac{Y_{A(K)}}{|K| !}~.
\label{curvedformalassign}
\eeq
The Heisenberg algebra \eq{heisalg} is traded for 
\beq
        [Y_{A(K)}, D_{i}  ] ~=~ |K|~ Y_{A(-i+ K)}~.      
\eeq
Remarkably, even in this non-commutative case, the exponentiated version can 
be recasted into the following simple form
\beq
     e^{q \ccdot Y} e^{D \ccdot y} ~=~ e^{(q+D)\ccdot y} e^{q\ccdot Y} ~.
\eeq
We shall have more to say about this construction at the end of
Appendix~\ref{appcalcorderedtuples}. The main point is that the proof
of the covariant Jacobi identity can be demonstrated in almost exactly 
the same way as in the flat metric case, cf.\  Appendix~\ref{noncomji}. 

\vs
\noi
Note that for non-zero curvature \mb{[D_{i},D_{j}]\phi^{A}(x,t)\neq 0},
the actual field value \mb{\phi^{A}(x,t)} is apparently not well-defined,
\ie if one tries to sum up the change in \mb{\phi^{A}(x,t)} along
a closed loop, one obtain a non-zero result. This is worse than a global 
obstruction. Perhaps it should be called a local obstruction. 
(A similar situation occures, say, in bosonic string theory with the 
Polyakov action when the worldsheet metric has a  non-zero curvature. 
This also leads to problems in locally assigning values to the target space 
fields.) The problem seems less formidable in the context of the Feynman path 
integral, where we only assign field values along one path at the time.
But it is a genuire challenge for the operator formalism.
One way of making sense out of this would be to 
declare the decendent fields \mb{\phi^{A(K)}(x,t)} to be independent 
fields living in a non-commutative jet-bundle.  In any case, we feel that 
it would be too hasty a priori to draw conclusions in general, and we leave 
it to the future to appropriately implement non-zero curvature in specific 
physical theories.


\setcounter{equation}{0}
\section{Supplementary Formalism}
\label{secsuppth}

\noi
Until now, we have only discussed functionals $F$ with no external space 
dependence, \ie all space-variables are integrated out. 
However for physical applications, we would like to conduct manipulations
directly on the integral kernels rather than the integrals.
For instance, to give sense to the fundamental equal-time relations
\beq
   \{ \phi^{A}(x,t),\phi^{B}(\yyy,t) \} ~=~\omega^{AB}(x,\yyy)~=~
\omega^{AB}~ \delta_{\Sigma}(x,\yyy) ~. 
\eeq 
The plan for the rest of this article are
\begin{itemize}
\item
to treat the Dirac delta distributions (and the derivatives thereof) in the 
presence of a boundary. Distributions is a vast subject in their own right, 
and we will here only give a heuristic treatment. 
\item
to extend the definition of the higher functional derivatives to more general 
types of local functionals.
\item
to analyse the implications for the Hamiltonian dynamics.
\end{itemize} 

\vs
\noi
While parts of this section is standard material, it
is reviewed for continuity and to fix notation.

\subsection{Embedded Approach}

\noi
Having a geometrically covariant formulation at our disposal enables
us to reduce the discussion to a single chart. We can slice up space in 
smaller regions; thereby producing unphysical double-sided boundaries 
(unphysical domain-walls), and we can hence consider space within
such a smaller region \mb{\Sigma} covered by a single chart. The local 
geometric data about the physical space \mb{\Sigma} is stored in the volume 
density \mb{\rho} and the  covariant derivatives \mb{D_{i}}. 

\vs
\noi
Furthermore, we will assume that \mb{\Sigma} takes place inside a 
{\em bounded} region of the chart \mb{\R^{d}}, \ie that it can be placed 
inside a large ball in \mb{\R^{d}}. Note that we are not placing any 
restriction on the distances in the physical space \mb{\Sigma}; only on 
the distances in the chart. Or perhaps we should say: in the choice of the 
chart. For instance, if space \mb{\Sigma=\R^{d}} is the ordenary flat space, 
one should map flat space into a bounded region \mb{\tilde{\Sigma}} 
of the chart \mb{\R^{d}} using a non-trivial \mb{\rho} and \mb{D_{i}}. 
In this case the spatial infinity is truly the boundary of the region 
\mb{\tilde{\Sigma}}. The perspective will be that of a typical Penrose 
diagram: ``There is always room for something beyond spatial infinity.'' 
The motivation for the above assumption is deeply founded in the theory of 
distributions, cf.\ below.

\vs
\noi
This being said, we will adapt the usual practice of 
identifying the space \mb{\Sigma \subseteq \R^{d}} with a region of the chart.

\subsection{Regularized Characteristic Functions}

\noi
Let us consider the characteristic function 
\beq
1_{\Sigma}(x)~=~\left\{\begin{array}{l} 1~~{\rm if}~~x\in\Sigma \cr
      0~~ {\rm otherwise}~, \end{array} \right.
\eeq
for the space \mb{\Sigma\subseteq \R^{d}} as a limit of a smooth function 
\mb{\chii_{\epsilon}(x)}, 
where \mb{0} \mb{<}  \mb{\chii_{\epsilon}(x)}\mb{\to}\mb{1_{\Sigma}(x)} for 
\mb{\epsilon \to 0^{+}}. We regard \mb{\chii_{\epsilon}=\chii_{\epsilon}(x)} 
as independent of the dynamical fields \mb{\phi^{A}(x)} and as a scalar
under coordinate changes \mb{x \to x'}. The actual implementation of the 
regularization \mb{\chii_{\epsilon}(x)} {\em should} not matter, so
one might as well choose a convenient form. 
One could for instance do as follows.  
Let  \mb{d_{\Sigma}(x)} denote the signed distance from \mb{x} 
to the boundary \mb{\pa \Sigma \subseteq \R^{d}} as measured in the chart 
\mb{\R^{d}}. The signed distance \mb{d_{\Sigma}(x)>0} is positive if 
\mb{x \in \Sigma^{\circ}} belongs to the interior and it is negative 
\mb{d_{\Sigma}(x)<0} if \mb{x \in (\R^{d} \backslash \Sigma)^{\circ}})
belongs to the exterior. Then we could implement \mb{\chii_{\epsilon}(x)} as
\beq
    \chii_{\epsilon}(x)~=~
\left(1+ \exp\left[- \frac{d_{\Sigma}(x)}{\epsilon}\right]\right)^{-1}~.
\eeq
(This looks horrible in other coordinates, so a geometrically minded
reader might prefer to substitute the chart \mb{\R^{d}} with an
abstract unphysical embedding manifold. We shall not explore this 
point of view further in this paper.)

\vs
\noi
Next, we extend \mb{\rho} and \mb{D_{i}} smoothly (and arbitrarily) 
to the unphysical sector \mb{\R^{d} \backslash \Sigma}. It may happen that 
\mb{\rho} or \mb{D_{i}} themselves are singular at the boundary 
\mb{\pa \Sigma}. In that case one should consider smooth regularized 
functions \mb{\rho_{\epsilon}} or \mb{D^{\epsilon}_{i}}, that in the limit
\mb{\epsilon \to 0} reproduces \mb{\rho} and \mb{D_{i}}.
Then an integral over \mb{\Sigma} should be though of as a limit 
\beq
\int_{\Sigma} \rho(x)~ d^{d}x~f(x) 
~=~ \lim_{\epsilon \to 0} 
\int \chii_{\epsilon}(x)\rho_{\epsilon}(x)~d^{d}x~ f_{\epsilon}(x)~. 
\label{inteps}
\eeq 
where the integrand \mb{f(x)} also should be smoothly (and arbitrarily) 
extended to the unphysical 
sector \mb{\R^{d} \backslash \Sigma}. (In case of more than one region, 
\mb{\chii_{\epsilon}(x)} should be a differentiable partition of the unity.)
{}From now on we will not write nor question the limit \mb{\epsilon \to 0}, 
but merely take for granted that this \mb{\epsilon}-prescription makes sense. 

\subsection{Dirac Delta Distributions}

\noi
Throughout the paper, the Dirac delta distribution 
\mb{\delta(x\!-\!z)} refers to the full \mb{\R^{d}}-chart, while
\mb{\delta_{\Sigma}(x,z)} refers to the physical space \mb{\Sigma}.
The physical Dirac delta distribution \mb{ \delta_{\Sigma}(x,z)} is
characterized by the property:
\beq
 \forall \eta~~:~~~~ \int_{\Sigma\times \Sigma} \rho(x)~ d^{d}x~ 
\rho(z)~ d^{d}z~\eta(x,z)~ \delta_{\Sigma}(x,z)
~=~ \int_{\Sigma} \rho(x)~ d^{d}x~ \eta(x,x)~.
\eeq
We can realize the physical Dirac delta distribution  
\mb{ \delta_{\Sigma}(x,z)} in terms of the unphysical Dirac delta 
distribution \mb{\delta(x\!-\!z)} as
\beq
   \delta_{\Sigma}(x,z)~=~
\frac{\delta(x\!-\!z)}{\chii_{\epsilon}(x)\rho(x)}~.
\label{deltarel}
\eeq
The main idea behind demanding that \mb{\Sigma} should occupy a {\em bounded}
region of the chart \mb{\R^{d}}, is that we can then perform formal 
integrations by part on the unphysical Dirac delta distributions 
\mb{\delta(x\!-\!z)}. This is so because we can consider all 
test functions as having a bounded support in the chart \mb{\R^{d}}. 
``Test functions'' should here be read in the broad sense of the word that 
in particular includes functions \mb{f(\phi^{(K)}(x),x)} of the dynamical 
fields \mb{\phi^{A(K)}(x)}.

\vs
\noi
On the other hand, integration by part of the 
physical Dirac delta distribution \mb{ \delta_{\Sigma}(x,z)} will in general 
lead to boundary contributions at the physical boundary \mb{\pa \Sigma}. 
The detailed form can be inferred from the above relation \eq{deltarel}.   
The benefit of this procedure, is that we do not have to postulate
perculiar rules for the physical delta distribution. They may simply be
derived from \eq{deltarel}.

\vs
\noi
The above is the key observation in our analysis of distributions. 
Mathematicians have always (and presumably for good reasons) considered 
test functions in \mb{\R^{d}} as having compact support. We observe that 
if space \mb{\Sigma}, which itself could be unbounded, fills a
bounded region of the chart \mb{\R^{d}}, we can without touching the 
above principle, still probe  boundary issues at the physical 
boundary \mb{\pa \Sigma}.

\vs
\noi
{}From \eq{inteps} and \eq{deltarel} it also becomes clear that the study of 
a non-trivial boundary \mb{\pa \Sigma} and the study of a non-trivial volume 
density \mb{\rho} are intimately related.
With the spatial integration interpreted as \eq{inteps}, we may define the
adjoint \mb{D_{i}^{\dagger}} of \mb{D_{i}} by formal integration by part
in the chart \mb{\R^{d}}. It becomes 
\beq
D_{i}^{\dagger}
~=~-\frac{1}{\chii_{\epsilon}\rho} \vec{D}_{i} 
\chii_{\epsilon}\rho (\cdot)~.
\eeq
(The arrow over \mb{D_{i}} indicates that the derivative \mb{D_{i}} acts 
{\em all} the way to the right.)

\subsection{Space of Functionals}

\noi
Consider now a function depending on variables \mb{z_{(1)},\ldots, z_{(r)}},
\beq 
  f(z_{(1)},\ldots, z_{(r)})~=~
f\left(D^{K_{(1)}}\phi(z_{(1)}),z_{(1)}, \ldots,
D^{K_{(r)}}\phi(z_{(r)}),z_{(r)} \right)~, 
\label{multifunc}
\eeq
where \mb{K_{(1)},\ldots, K_{(r)}} are multi-indices. 
{}For convenience we shall often use the compact 
notation \mb{z\equiv(z_{(1)},\ldots, z_{(r)})} if there is several 
space points \mb{z_{(i)}}.
We shall restrict ourselves to the space \mb{{\cal A}} of functionals $F$ 
that can be expressed as a $s$-fold multiple integral over \mb{\Sigma}
\bea
   F(z)~=~\int_{\Sigma \times \ldots \times \Sigma} 
\rho d^{d}z_{(1)} \ldots \rho d^{d}z_{(s)}~  f(z)~,
\label{lmultfuncal}
\eea  
for some \mb{s \in \N}. (It is implicitly understood that the \mb{z_{(i)}}'s 
which are integrated out on the right hand side, do not enter the argument 
list on the left hand side.)
Furthermore, this space \mb{{\cal A}}  is a \mb{\C}-vector space. 
It is stabile under multiplication whenever defined. (Recall that
the product of distributions need not be well-defined.) It is closed
under integrating out external variables, or identifying external variables, 
say \mb{z_{(i)}=z_{(j)}}. 
We shall see below that it is also closed under the full Poisson bracket.

\subsection{Suitable Form of Functional}
\label{secsuitform}

\noi
Consider a local functional 
\mb{F(z)=\int d^{d}x_{(1)} \ldots d^{d}x_{(r)}~f(x,z)} with external 
dependence \mb{z}. The typical integrand \mb{f(x,z)} consists of 
\begin{enumerate}
\item
The Dirac delta distributions \mb{\delta(x\!-\!\yyy)} and 
\mb{\delta_{\Sigma}(x,\yyy)}; The regularized characteristic functions 
\mb{\chii^{p}_{\epsilon}(x)} in some  power  \mb{p\in \R}.
\item
Smooth test functions with bounded support. This in particular includes 
smooth functions \mb{g(\phi^{(K)}(x),x)} of the dynamical fields 
\mb{\phi^{A(K)}(x)} and the smooth volume density \mb{\rho(x)}.
\item Derivatives \mb{D_{i}} acting on the various factors of the integrand
mentioned under point \mb{1-2}.  
\end{enumerate}
The above listed objects appears in two versions:
\begin{itemize}
\item[\mb{A}.] an {\em external} type, if it depends on non-integrated 
external 
variables.
\item[\mb{B}.] an {\em internal} type, if it (at least partially) 
depends on integrated internal variables.
\end{itemize}
An integral (kernel) is declared to be on {\em suitable} form if all 
internal derivatives \mb{(B3)} act on type \mb{B2} objects. In 
other words that the more singular type \mb{(B1)} objects are not hit
by the the internal derivatives \mb{(B3)}.

\vs
\noi
A functional (kernel) \mb{f(x,z)} is not well-defined if one cannot
obtain a suitable form by pure algebraic manipulations. 
In practice, this means
\begin{itemize}
\item
after formal internal integration by part, 
\item
after use of the Leibnitz rule and linarity.
\item 
after use of the identity
\mb{\left(g(x)-g(y)\right)\delta(x\!-\!\yyy)~=~0}~,
\item 
and after use of the identity
\mb{\left(D_{(x)}+D_{(y)}\right)\delta(x\!-\!\yyy)~=~0}~
\end{itemize}
for the (partially) internal variables.

\vs
\noi
Just from the freedom to perform a formal integration by part on the
internal differentiation \mb{(B3)} (or choosing not to do so, respectively),  
there is
\mb{2^{n}} ways of writing down an functional, where \mb{n} is the number 
of internal differentiations \mb{(B3)}. In practice, in all interesting 
functionals, every internal variable \mb{x_{i}} appear at least once in 
the argument list of an internal singular object (\ie of type \mb{B1}). 
As a consequence, in these cases, at most one of the above mentioned 
\mb{2^{n}} choices leads to a suitable form.

\vs
\noi
Needless to say that if one also integrate out the external \mb{z}-variables 
\mb{F(z)} without yielding enough room for the smearing type \mb{2} objects,
the result may not be well-defined.

\subsection{Higher Partial Derivatives}

\noi
Consider a functional \mb{F(z) \in {\cal A}} in the algebra \mb{{\cal A}}.
Assume that the functional (kernel) is of suitable form.
Let us now define the higher partial derivatives as
\beq
\papatwo{F(z)}{\phi^{A(K)}(x)}~\equiv~
\int_{\Sigma^{s}}   \rho d^{d}z_{(1)} \ldots \rho d^{d}z_{(s)}    
\sum_{i=1}^{r} 
\delta_{\Sigma}(x,z_{(i)})~P_{A(K)}^{(z_{(i)})} f(z)~.
\label{highpartintfuncal}
\eeq
In case of a {\em function} 
\beq
F(z)~=~F\left(D^{K}\phi(z),z\right)~,
\eeq
this reduces to
\beq
\papatwo{F(z)}{\phi^{A}(x)}~=~
 \sum_{i=1}^{r} 
\delta_{\Sigma}(x,z_{(i)})~P_{A(K)}^{(z_{(i)})} F(z)~.
\label{highpartintfunc}
\eeq
We can formally extend the application range of the above equation 
\eq{highpartintfunc} to include functionals \mb{F(z)} as well, by implicitly 
assuming that the internal delta distributions automatically are placed 
inside the integration symbol. Then \eq{highpartintfunc} becomes a 
convenient shorthand notation for \eq{highpartintfuncal}.

\subsection{Higher Functional Derivatives}

\noi
The general definition \eq{ordhighderiv} for a functional of suitable form 
yields
\bea
\dedetwo{F(z)}{\phi^{A(K)}(x)}
&=&E_{A(K)}^{(x)}\sum_{i=1}^{r}\left[\delta_{\Sigma}(x,z_{(i)})~
F\left(z_{(1)},\ldots, z_{(i-1)},x,z_{(i+1)},\ldots \right) \right] \cr
&=&\sum_{i=1}^{r}\sum_{M \succeq K} (-D_{(x)})^{M^{t}\div K^{t}} 
\delta_{\Sigma}(x,z_{(i)})~P_{A(M)} ^{(z_{(i)})}F(z)~.
\label{highfuncintfunc}
\eea
The above derivatives of a delta distribution may be resolved in two 
different ways: 
\begin{itemize}
\item
By {\em inner} evaluation: 
The derivatives leave the delta distribution \mb{\delta_{\Sigma}(x,z_{(i)})} 
via the \mb{z_{(i)}}-leg.
If there is enough internal integrations inside the functional \mb{F}, 
one may resolve the delta distributions by integration, thereby 
prolonging the derivative to an object inside the functional \mb{F}. 
If all terms are to be resolved this way, this means that all the 
\mb{z}-variables have to be internal.
\item
By {\em outer} evaluation: The derivatives leave the delta distribution 
\mb{\delta_{\Sigma}(x,z_{(i)})} via the \mb{x}-leg.
We await an external integration over the \mb{x}-variable 
to evaluate the derivative of the delta distribution by formal integration 
by part. Let us stress the fact, that if one rely on the latter 
method for the \mb{\delta_{\Sigma}(x,z_{(i)})} term in \eq{highfuncintfunc} 
with the \mb{z_{(i)}} being an internal variable for the \mb{F(z)} functional, 
then the \mb{x}-integrated version is not on a suitable form as it stands. 
\end{itemize}
This distinction is important if one is to conduct further partial 
differentiations  wrt.\ to the dynamical fields \mb{\phi^{A}(x)} on the 
functional (kernel). However, if no further differentiations are performed,
the two methods yields the same result.

\vs
\noi   
As the most important example, we mention
\beq
\papatwo{\phi^{B}(\yyy)}{\phi^{A}(x)}
~=~\dedetwo{\phi^{B}(\yyy)}{\phi^{A}(x)}
~=~\delta^{B}_{A}~ \delta_{\Sigma}(x,\yyy)~.
\eeq
Note that the above definitions \eq{highpartintfunc} and \eq{highfuncintfunc}
guarantee the linearity and 
the Leibnitz' rule of the (higher) partial and functional derivatives. 
We also find that two (higher) partial derivatives commute.
One may show in the case of a vanishing boundary 
\mb{\pa \Sigma = \emptyset}, 
that the usual functional derivatives commute. In the case of a non-trivial 
boundary \mb{\pa \Sigma\neq\emptyset}, the higher functional derivatives
(as well as the usual functional derivatives) do {\em not} commute
in general.

\subsection{Induced Functional Derivative}

\noi
Finally, one may define a induced functional 
derivative from the perspective of the embedding manifold, \ie the chart 
\mb{\R^{d}}:
\beq
\delta F(z)~=~\int d^{d}x~
\dedetwo{[\chii_{\epsilon}(x)\rho(x)F(z)]}{\phi^{A}(x)}\delta \phi^{A}(x)~,
\label{traddiffrevisited}
\eeq  
for an arbitrary variation \mb{\delta \phi^{A}(x)}. 
This is so, because \mb{\Sigma\subseteq \R^{d}} is bounded 
inside the chart \mb{\R^{d}}, so integration by part yields no boundary 
contributions at \mb{|x|=\infty}. The induced functional derivative 
makes sense, eventhough there appears coinciding space points, 
because \mb{\chii_{\epsilon}(x)\rho(x)} does not depend on the dynamical 
fields \mb{\phi^{A}(x)}. We can write it constructively as
\beq
\dedetwo{[\chii_{\epsilon}(x)\rho(x) F(z)]}{\phi^{A}(x)}
~=~\sum_{i=1}^{r}\sum_{M \succeq \emptyset} (-D_{(x)})^{M^{t}} 
\delta(x\!-\!z_{(i)})~P_{A(M)} ^{(z_{(i)})}F(z)~.
\label{chihighfuncintfunc}
\eeq
It is related to the higher functional derivatives via
\beq
\dedetwo{[\chii_{\epsilon}(x)\rho(x)  F(z)]}{\phi^{A}(x)}
~=~(-D_{(x)})^{M^{t}}\left(\chii_{\epsilon}(x)\rho(x)\right)
\dedetwo{F(z)}{\phi^{A(M)}(x)}~.
\label{funcderrevisited}
\eeq
This induced functional derivative has the remarkable property of commuting 
with the spatial derivatives
\beq
D_{i}^{(z_{(j)})} \dedetwo{[\chii_{\epsilon}(x)\rho(x) F(z)]}{\phi^{A}(x)}
~=~\dedetwo{[\chii_{\epsilon}(x)\rho(x) 
D_{i}^{(z_{(j)})}F(z)]}{\phi^{A}(x)}~.
\label{niceind1}
\eeq
This should be compared with the corresponding behaviour of the usual
functional derivatives:
\beq
-D_{i}^{\dagger(z_{(j)})} \dedetwo{ F(z)}{\phi^{A}(x)}
~=~\dedetwo{[ D_{i}^{(z_{(j)})}F(z)]}{\phi^{A}(x)}~.
\label{notsonice1}
\eeq
The induced functional derivative  satisfies the Leibnitz rule and it 
commutes with integrations:
\beq
  \int_{\Sigma}  \rho d^{d}z_{(i)}~ 
\dedetwo{[\chii_{\epsilon}(x)\rho(x) F(z)]}{\phi^{A}(x)}
~=~\dedetwo{\left[\chii_{\epsilon}(x)\rho(x)  
\int_{\Sigma} \rho d^{d}z_{(i)}F(z)\right]}{\phi^{A}(x)}~.
\label{niceind2}
\eeq

\subsection{Annihilation Principle}

\noi
As mentioned before, integration without smearing can produce ill-defined
terms. However it can be very cumbersome to a priori discard all the
bad terms of an expression. We shall therefore formally allow ill-defined
terms to appear by giving a prescription, that consistently identify them and 
put them to zero. This is done by defining a little more restrictive
version of the above so-called suitable form.
{\em Notation: For simplicity, we will assume from now on that the covariant 
derivatives commute, \ie that the curvature vanishes. Assume further from 
now on that the volume density \mb{\rho} is covariantly preserved, 
\mb{D_{i}\rho=0}.} 
A typical functional (kernel) consists of
\begin{enumerate}
\item
Dirac delta distributions \mb{\delta(x\!-\!\yyy)}. 
\item
Regularized characteristic functions \mb{\chii_{\epsilon}(x)}. (Integral 
powers \mb{\chii_{\epsilon}^{n}(x)}, \mb{n\geq 2} should be considered 
as a $n$-fold product of elementary \mb{\chii_{\epsilon}(x)}.)
\item
Negative powers \mb{\chii_{\epsilon}^{-p}(x)}, \mb{p \geq 0}, of the 
regularized characteristic function.
\item
Smooth test functions with bounded support. This includes smooth functions 
\mb{g(\phi^{(K)}(x),x)} of the dynamical fields \mb{\phi^{A(K)}(x)} and
the smooth volume density \mb{\rho(x)}.
\item Derivatives \mb{D_{i}} acting on various factors of the integrand
mentioned under point \mb{1-4}.  
\end{enumerate}
We may assume by use of the Leibnitz rule and breaking the integral 
\mb{F(z)} into several terms if necessary, that all derivatives 
\mb{(A5+B5)} only act on one elementary object under point \mb{1}, 
\mb{2} and \mb{4}. (Here the letters \mb{A} and \mb{B} refers to the 
notation introduced in Subsection~\ref{secsuitform}, while the numbers 
\mb{1-5} are those defined above in this subsection.)

\vs
\noi
A functional \mb{F(z)} of the above atomic type is declared to be
identical zero if one cannot by algebraic means, cf.\  above,  
obtain a form where all internal derivatives \mb{(B5)}
acts on type \mb{B4} objects. (Or in other words,  the more singular
type of objects \mb{B1-B3} are not hit by the derivatives.)

\setcounter{equation}{0}
\section{Hamiltonian Edge Dynamics}
\label{seched}

\noi
In this section we discuss implications of the full Poisson bracket for the
Hamiltonian dynamics. We first have to extend the definition \eq{fullpb2}
of the full Poisson bracket to more general functionals with external 
dependence. As a first principle for writing down the more general Poisson 
bracket, we shall demand that
integrations \mb{\int_{\Sigma} \rho d^{d}z_{(i)}} commute with
the Poisson bracket \mb{\{ \cdot , \cdot \}}, that is
\beq
\int_{\Sigma} \rho d^{d}z_{(i)}~\{F(z),G(w)\}
~=~ \{ \int_{\Sigma} \rho d^{d}z_{(i)}~ F(z),G(w)\}~.
\eeq
This principle leads naturally (modulo the action of the annihilation 
principle) to what we shall call the {\em solid} Poisson bracket.  We shall 
later see that it can be recasted into a so-called {\em floating} Poisson 
bracket, that (at a superficial level) takes different shape on different 
types of functionals. However, one may show by applying the annihilation 
principle that no actual differences take place.

\subsection{Solid Poisson Bracket}

\noi
Using the extrapolation of the formulas in the previous sections, the  
full Poisson bracket becomes
\beq
  \{ F(z), G(w) \}~=~\{F(z) , G(w) \}_{(0)} + B(F(z),G(w)) ~,
\eeq
where
\bea
 \{ F(z), G(w) \}_{(0)} 
&=& \int_{\Sigma} \rho(x) d^{d}x 
\dedetwo{F(z)}{\phi^{A}(x)} \omega^{AB}  \dedetwo{G(w)}{\phi^{B}(x)}  \cr
&=&\int_{\Sigma \times \Sigma} \rho(x) d^{d}x~ \rho(\yyy) d^{d}\yyy~
\dedetwo{F(z)}{\phi^{A}(x)} \omega^{AB}(x,\yyy) 
\dedetwo{G(w)}{\phi^{B}(\yyy)}  ~, \cr
B(F(z),G(w))&=& \sum_{K\neq \emptyset} \int_{\Sigma} \rho(x) d^{d}x ~
D_{(x)}^{K} \left[  
\dedetwo{F(z)}{\phi^{A(K)}(x)} \omega^{AB} 
\dedetwo{G(w)}{\phi^{B}(x)} \right]~-~ (F(z) \leftrightarrow G(w)) \cr
&=& \sum_{K\neq \emptyset} \int_{\Sigma} \rho(x) d^{d}x ~
 \left[(D^{\dagger}_{(x)})^{K^{t}} 
\papatwo{F(z)}{\phi^{A(K)}(x)}\right] \omega^{AB} 
\dedetwo{G(w)}{\phi^{B}(x)} \cr 
&&~-~ (F(z) \leftrightarrow G(w))~.
\label{fullpb3}
\eea  
Alternatively, we may write the full Poisson bracket as
\beq
\{ F(z), G(w) \}
~=~\int_{\Sigma \times \Sigma} \rho(x) d^{d}x~ \rho(\yyy) d^{d}\yyy~
\papatwo{F(z)}{\phi^{A(M)}(x)}\omega^{A(M)B(N)}(x,\yyy) 
\papatwo{G(w)}{\phi^{B(N)}(\yyy)} ~,
\eeq
where the symplectic kernel \mb{\omega^{A(M)B(N)}(x,\yyy)} reads
\bea
\omega^{A(M)B(N)}(x,\yyy)
&=&  \omega^{AB}  \left[ 
- (- D^{\dagger}_{(x)})^{M}  (-D^{\dagger}_{(\yyy)})^{N} \right. \cr
&&+  \left. (- D^{\dagger}_{(x)})^{M}  (D^{\dagger}_{(x)})^{N^{t}}
+(- D^{\dagger}_{(\yyy)})^{N}  (D^{\dagger}_{(\yyy)})^{M^{t}}
 \right]  \delta_{\Sigma}(x,\yyy) \cr
&=&  \omega^{AB}  \left[ 
- ( -D^{\dagger}_{(x)})^{M}  (-D^{\dagger}_{(\yyy)})^{N}   \right.  \cr
&&+  \left.( -D^{\dagger}_{(x)})^{M}  D_{(\yyy)}^{N}
+ D_{(x)}^{M}  (-D^{\dagger}_{(\yyy)})^{N}
 \right]  \delta_{\Sigma}(x,\yyy) ~.
\label{symplkernel}
\eea
In the case where (at least) one of the \mb{M} and \mb{N} 
are \mb{\emptyset}, the symplectic kernel can neatly be written as
\beq
M~=~\emptyset~ \vee~ N~=~\emptyset~~~~~~~\Rightarrow~~~~~~~
\omega^{A(M)B(N)}(x,\yyy)
~=~\omega^{AB}~ D_{(x)}^{M}  D_{(\yyy)}^{N} \delta_{\Sigma}(x,\yyy) ~.
\eeq
Note on the other hand, that the case \mb{M=\emptyset \vee N=\emptyset} 
is also the maximal case to make fully sense out of the expression
\mb{D_{(x)}^{M}  D_{(\yyy)}^{N} \delta_{\Sigma}(x,\yyy)} without
employing the annihilation principle. Beyond that case, \ie if
\mb{M\neq\emptyset \wedge N\neq\emptyset}, there is no
escape ways left open for the \mb{\chii^{-1}_{\epsilon}}-function (via 
formal integrations by part of the derivatives). It is sandwiched inside 
the delta distribution between derivatives.
It should be merged with a \mb{\chii_{\epsilon}}-function outside, whose mere
existence on the other hand prohibes that a suitable (and hence a 
well-defined) form can be reached by means of integration by part.

\subsection{Hamilton Equations of Motion} 

\noi
Consider the Hamilton equations of motion
\beq
    {{d} \over {dt}} F(z) 
~=~-\{H, F(z) \} + {{\pa} \over {\pa t}} F(z)~.
\label{hameqofmotion}
\eeq
Let the Poisson bracket be the solid Poisson bracket \eq{fullpb3}.
And \mb{H=\int_{\Sigma} \rho(x) d^{d}x~{\cal H}(x)} be a local Hamiltonian.
Then the Hamilton equations of motion for the fundamental fields read
\beq
{{d} \over {dt}} \phi^{A}~=~\{ \phi^{A}, H \}
~=~  \frac{\omega^{AB}}{\chii_{\epsilon} {}} 
E_{B(0)}\left(\chii_{\epsilon}{} {\cal H} \right)~.
\label{hmfundhameqmot}
\eeq
We also get 
\beq
{{d} \over {dt}} \phi^{A(K)}~=~\{ \phi^{A(K)}, H \}
~=~ \frac{\omega^{AB}}{\chii_{\epsilon} {}}   \left( D^{K}
E_{B(0)}\left(\chii_{\epsilon} {} {\cal H}  \right)
-\left[ D^{K}, \chii_{\epsilon}{} \right]  E_{B(0)} {\cal H} \right) ~.
\label{hmm}
\eeq
On the other hand, a spatial differentiation of \eq{hmfundhameqmot} yields
\beq
D^{K}{{d} \over {dt}} \phi^{A}~=~D^{K} \{ \phi^{A}, H \}
~=~ \omega^{AB} D^{K} \frac{1}{\chii_{\epsilon} {} } 
E_{B(0)}\left(\chii_{\epsilon}{} {\cal H}  \right)~.
\label{hmmm}
\eeq
Superficially, the spatial differentiations do not commute with the time 
derivatives; compare \eq{hmm} and \eq{hmmm}. But we shall see that this is
an illusion. First of all, it is easy to localize potential problems to 
the boundary \mb{\pa \Sigma}. Away from the boundary, the two expressions
\eq{hmm} and \eq{hmmm} fully agree. However, being interested in the 
boundary dynamics, this does not quite satisfy us.
Let us check that they also agree on the boundary. Clearly, the expressions
are singular at the boundary, so the only way to extract meaningful
information is to prepare both the expressions with a general test function 
\mb{\eta_{A}(x)}.  It is not difficult to see that the annihilation 
principle sweeps away any differences between the two smeared expressions:
\beq
\int_{\Sigma} \rho(x) d^{d}x~\eta_{A}(x)~
[{{d} \over {dt}} , D^{K}]\phi^{A}(x)~=~0~.  
\eeq

\subsection{Floating Poisson bracket}

\noi
The floating Poisson bracket is defined via the induced functional 
derivatives
\bea
\{ F(z), G(w) \}
&=&\int_{\Sigma} \rho d^{d}x~\chii_{\epsilon}^{-1}(x) 
\dedetwo{[\chii_{\epsilon}(x) F(z)]}{\phi^{A}(x)}\omega^{AB} 
\chii_{\epsilon}^{-1}(x)\dedetwo{[\chii_{\epsilon}(x) G(w)]}{\phi^{B}(x)} \cr
&=&\int_{\Sigma} \rho d^{d}x~ 
\frac{(-D)^{M^{t}}\chii_{\epsilon}(x)}{\chii_{\epsilon}(x)} 
\dedetwo{F(z)}{\phi^{A(M)}(x)}\omega^{AB}
\frac{(-D)^{N^{t}}\chii_{\epsilon}(x)}{\chii_{\epsilon}(x)} 
\dedetwo{G(w)}{\phi^{B(N)}(x)}  \cr
&=&\int_{\Sigma \times \Sigma} \rho d^{d}x~ \rho d^{d}\yyy~
\papatwo{F(z)}{\phi^{A(M)}(x)}  \left[ D_{(x)}^{M}  D_{(\yyy)}^{N} 
\delta_{\Sigma}(x,\yyy) \right]\omega^{AB}\papatwo{G(w)}{\phi^{B(N)}(\yyy)}~.
\label{fullpb4}
\eea
In the second equality, we used the identity \eq{funcderrevisited}. Let us 
check that the floating Poisson bracket \eq{fullpb4} becomes the full Poisson 
bracket \eq{fullpb2}, by use of the annihilation principle when \mb{F} and 
\mb{G} are both local functionals with no external dependence. This is 
perhaps best seen from the second right hand side of \eq{fullpb4}. 
The idea is now to obtain a restrictive suitable form by recasting all the 
derivatives onto smooth objects. In the case at hand, the higher functional 
derivatives are both smooth functions of \mb{x}. One may deduce that the
required form can only be obtained for terms \mb{(M,N)} when at least one 
of the indices \mb{M}, \mb{N} are \mb{\emptyset}.
This reproduces precisely the full Poisson bracket \eq{fullpb2}.
Technically speaking, for more general functionals \mb{F(z)} and \mb{G(w)}, 
the above truncation takes place on parts of the functionals where both
functionals are evaluated by the inner method. If at least one of them
is evaluated by the outer method, then no truncation is carried out (although
a truncation may take place at a later integration).
This is particular the case for two functions  \mb{F(z)} and \mb{G(w)}.

\vs
\noi
The advantage of the floating formulation is at least two-fold: First of all,
the full Poisson bracket can be written in a more compact manner. 
The second  reason is that the floating Poisson bracket manifestly commutes 
with the spatial derivatives, cf.\ \eq{niceind1}. This is quite useful.
For instance, consider as previously, the Hamilton equations of motions,
but this time with the floating  bracket as the Poisson bracket. 
We get as before
\beq
{{d} \over {dt}} \phi^{A}~=~\{ \phi^{A}, H \}
~=~  \frac{\omega^{AB}}{\chii_{\epsilon}} 
E_{B(0)}\left(\chii_{\epsilon}{\cal H} \right)~.
\label{hmfundhameqmot2}
\eeq
However, this time the spatial derivatives 
commute manifestly with the time evolution
\beq
[ {{d} \over {dt}} , D^{K} ] \phi^{A}~=~
\{  \phi^{A(K)}, H \} -D^{K} \{ \phi^{A}, H \}  ~=~0~.
\label{hmmm2}
\eeq
At the end of the day, it always boils down to the full Poisson bracket 
\eq{fullpb2}, when all the variables are integrated out (and the 
annihilation principle applied). We voluntarily throw in a lot of formal 
zeroes in the floating Poison bracket dressed up in the above 
``divergent-looking''  disguise. This gambit enables us to write the full 
Poisson bracket in a very compact manner. 

\vs
\noi
As perhaps the most important point, let us note that the Hamilton 
equations of motions \eq{hmfundhameqmot2} follows from extremizing, 
in the sense of \eq{traddiffrevisited}, the following natural action:
\beq
 S~=~ \int dt \int_{\Sigma} \rho d^{d}x \left[ \frac{1}{2}\phi^{A}(x,t) 
\omega_{AB} \dot{\phi}^{A}(x,t) - {\cal H}(x,t) \right]~. 
\eeq

\vs
\noi
One may define, a more general floating \mb{\alpha}-bracket, 
for \mb{\alpha \geq 1}, where   
\bea
\{ F(z), G(w) \}_{(\alpha)}
&=&\int_{\Sigma} \rho d^{d}x~\chii_{\epsilon}^{-1}(x) 
\dedetwo{[\chii^{\alpha}_{\epsilon}(x) F(z)]}{\phi^{A}(x)}\omega^{AB} 
\chii_{\epsilon}^{-1}(x)
\dedetwo{[\chii^{\alpha}_{\epsilon}(x) G(w)]}{\phi^{B}(x)} \cr
&=&\int_{\Sigma} \rho d^{d}x~ 
\frac{(-D)^{M^{t}}\chii^{\alpha}_{\epsilon}(x)}{\chii_{\epsilon}(x)} 
\dedetwo{F(z)}{\phi^{A(M)}(x)}\omega^{AB}
\frac{(-D)^{N^{t}}\chii^{\alpha}_{\epsilon}(x)}{\chii_{\epsilon}(x)} 
\dedetwo{G(w)}{\phi^{B(N)}(x)} ~.
\label{fullpb4a}
\eea
In this paper, we will merely view the floating \mb{(\alpha\!>\!1)}-brackets 
as a curiosity, which nevertheless has some relevance  when addressing 
time-order issues, see next Section~\ref{timeord}.
It coincide in the inner sector with the above floating 
\mb{(\alpha\!=\!1)}-bracket. And it satisfies the Jacobi identity. We give an
independent proof of the important
the \mb{\alpha=1} case in the Appendix~\ref{singji}, and leave the general
case \mb{\alpha > 1} to the reader.
The \mb{\alpha}-factor slows down the \mb{\epsilon}-convergence process, 
but viewed as an isolated issue, it does not jeopardize the convergence.

\section{Time Order and Quantization}
\label{timeord}

\noi
One might suspect that the boundary terms are related to the time order
prescriptions. We can give some heuristic arguments which points in 
that direction. 
Consider first the following totally ordered, antisymmetric and transitive 
time order prescription for two spacetime points \mb{(x_{(1)},t_{(1)})} 
and \mb{(x_{(2)},t_{(2)})} in unphysical space-time \mb{\R^{d} \times \R}.
\bea
  (x_{(1)}, t_{(1)})~\prec~(x_{(2)}, t_{(2)})&\Leftrightarrow&
\left(x_{(1)}, x_{(2)}~ \in~  \Sigma~ \wedge~ 
t_{(1)}~<~ t_{(2)} \right)\vee~\left(
x_{(1)}~ \notin~  \Sigma~ \wedge~  x_{(2)}~ \in~  \Sigma \right)\cr
&&\vee~\left(
x_{(1)}, x_{(2)} ~\notin~ \Sigma ~ \wedge~  
|d_{\Sigma}(x_{(1)})|~>~|d_{\Sigma}(x_{(2)})| \right)~.
\eea
This prescription has as a consequence, that the boundary \mb{\pa\Sigma}  
is assigned to the infinite past, so there effectively is no spatial 
boundary. (In a similar manner, one may link it with the infinite future.) 
As we shall see below, the equal-time relation \mb{\sim} plays a 
crucial role, so let us define it properly:
\bea
  (x_{(1)}, t_{(1)})~\sim~(x_{(2)}, t_{(2)})&\Leftrightarrow&
\left(x_{(1)}, x_{(2)}~ \in~  \Sigma~ \wedge~ 
t_{(1)}~ =~ t_{(2)} \right)\cr
&&\vee~\left(
x_{(1)}, x_{(2)} ~\notin~ \Sigma ~ \wedge~  
|d_{\Sigma}(x_{(1)})|~=~|d_{\Sigma}(x_{(2)})| \right)~.
\eea
We can now give the time order prescription \mb{T_{\Sigma}} for \mb{n} 
operators \mb{\hat{F}_{(i)}=\hat{F}_{(i)}(x_{(i)}, t_{(i)})}, where
\mb{i=1,\ldots,n}.
\beq
T_{\Sigma}\left[ \hat{F}_{(n)}\ldots \hat{F}_{(1)} \right]
~=~\hat{F}_{(\pi(n))}\ldots \hat{F}_{(\pi(1))}
\eeq
where \mb{\pi \in S_{n}} is the unique permutation
\mb{\{1,\ldots,n\} \to \{1,\ldots,n\}}, that satisfies
\bea
 (x_{(n)}, t_{(n)}) ~\succeqq &\ldots& \succeqq ~(x_{(1)}, t_{(1)})~, \cr \cr
\forall i=1,\ldots,n-1:~ \pi(i\!+\!1) ~<~  \pi(i) &\Rightarrow& 
 (x_{(\pi(i\!+\!1))}, t_{(\pi(i\!+\!1))})~ 
\not\sim~ (x_{(\pi(i))}, t_{(\pi(i))})~.
\eea
Let us now consider an equal-time slice \mb{t=t_{(0)}}, \ie in the 
traditional sense of the word ``equal-time'',  as we did in the previous 
sections. We shall suppress the time coordinate in the following. We have
\bea
 T_{\Sigma}[ \hat{F}(x), \hat{G}(y)]
&=&\left[ 1_{\Sigma}(x)~1_{\Sigma}(y) 
+ 1_{\{0\}}\left(  |d_{\Sigma}(x_{(1)})|-|d_{\Sigma}(x_{(2)})| \right)
\left(1-1_{\Sigma} (x) \right)\left(1- 1_{\Sigma}(y) \right)\right]
[ \hat{F}(x), \hat{G}(y)] \cr
&\approx& 
 1_{\Sigma}(x)~ 1_{\Sigma}(y)~[ \hat{F}(x), \hat{G}(y)]~. 
\eea  
In the wavy equality \mb{\approx}, we neglected a contribution from a 
spatial hypersurface of dimension \mb{d-1}, and hence of Lebesgue measure 
zero. Let us regularized this as
\beq
T^{\beta}_{\Sigma}[ \hat{F}(x), \hat{G}(y)]~=~
\chii^{\beta}_{\epsilon}(x)~\chii^{\beta}_{\epsilon}(x)~
[ \hat{F}(x), \hat{G}(y)]~,
\eeq
for some positive power \mb{\beta >0}.
When time ordering the spatial derivatives we get
\beq
D_{(x)}^{M} D_{(y)}^{N} T^{\beta}_{\Sigma}[ \hat{F}(x), \hat{G}(y)]
~=~\frac{1}{\chii^{\beta}_{\epsilon}(x)}~
\frac{1}{\chii^{\beta}_{\epsilon}(y)}~
T^{\beta}_{\Sigma}\left[ D_{(x)}^{M} 
\left(\chii^{\beta}_{\epsilon}(x)\hat{F}(x)\right),
D_{(y)}^{N}\left( \chii^{\beta}_{\epsilon}(y)\hat{G}(y)\right) \right]~.
\eeq
(This equation should be understood as follows: On the left hand side
the time order \mb{T^{\beta}_{\Sigma}} is everywhere present while we form 
the quotient of differences for the derivative \mb{D_{i}}. In particular, it
is present before we actually take the infinitesimal 
limit to produce the derivatives \mb{D_{i}}. On the right hand side 
the time order \mb{T^{\beta}_{\Sigma}} does not recognize how the spatial 
derivatives earlier were produced. \mb{T^{\beta}_{\Sigma}} only sees the 
result: an operator depending on one space point.)
This should be compared to the corresponding property of the floating
\mb{(\alpha\!>\!1)}-Poisson bracket
\beq
D_{(x)}^{M} D_{(y)}^{N} \left\{ \hat{F}(x), \hat{G}(y) \right\}_{(\alpha)}
~=~\frac{1}{\chii^{\alpha\!-\!1}_{\epsilon}(x)}~
\frac{1}{\chii^{\alpha\!-\!1}_{\epsilon}(y)}~\left\{ D_{(x)}^{M} 
\left(\chii^{\alpha\!-\!1}_{\epsilon}(x)\hat{F}(x)\right),
D_{(y)}^{N}\left( \chii^{\alpha\!-\!1}_{\epsilon}(y)\hat{G}(y)\right) 
\right\}_{(\alpha)}~.
\eeq
This carries some evidence, that we should translate the floating
\mb{(\alpha\!>\!1)}-Poisson bracket into the commutator with the 
above perculiar time order prescription, with \mb{\beta=\alpha-1}:
\beq
\frac{1}{i \hbar} T^{\beta=\alpha\!-\!1}_{\Sigma}[ \hat{F}(x), 
\hat{G}(y)] ~~~\leftrightarrow~~~ \{F(x) , G(y)\}_{(\alpha)}~.
\eeq
Although the exact value of \mb{\alpha>1} should be taken with a grain of 
salt, let us compare this behavior with the behavior
of the floating \mb{(\alpha\!=\!1)}-Poisson bracket:
\beq
D_{(x)}^{M} D_{(y)}^{N} \{F(x) , G(y)\}
~=~
\left\{ D_{(x)}^{M} F(x),
D_{(y)}^{N} G(y) \right\}~,
\eeq
This corresponds to the commutator with the usual time order prescription. 
It is also interesting to compare with the corresponding property of the 
bulk Poisson bracket, cf.\ \eq{notsonice1}, although it of course does 
not satisfy the Jacobi identity and is therefore not expected to play any 
leading role at the level of quantization:
\beq
\frac{1}{\chii_{\epsilon}(x)}~\frac{1}{\chii_{\epsilon}(y)}~
D_{(x)}^{M} D_{(y)}^{N} \left\{\chii_{\epsilon}(x) F(x) , 
\chii_{\epsilon}(y) G(y)\right\}_{(0)}
~=~\left\{ D_{(x)}^{M}F(x), D_{(y)}^{N}G(y) \right\}_{(0)}~.
\eeq

\vs
\noi
So we have here presented two physically different, but both consistent, 
time orderings. One governed by the floating \mb{(\alpha\!>\!1)}-Poisson 
bracket, but with some of the ``equal-time'' surfaces wrapped up along the 
boundary \mb{\pa \Sigma}. And another system governed by the floating 
\mb{(\alpha\!=\!1)}-Poisson bracket with the boundary 
\mb{\pa \Sigma} being a true spatial boundary for the system. Although the
above analysis clearly may be criticized wrt.\ 1) the order of various limits
taken, 2) the omission of the role played by the annihilation principle, and 
3) its disregards of further ordering issues (like \mb{*}-product), 
it tends to confirm the importancy of the boundary terms 
of the Poisson bracket, and that they should not be
discarded in a full treatment of a quantum field theory with a spatial 
boundary.


\setcounter{equation}{0}
\section{Conclusions}

\noi
In this article we have
\begin{itemize}
\item
Reviewed the higher functional derivatives and the extended notion 
of differentiability of functionals. 
\item
Shown a new way to add a boundary contribution to the 
usual ``bulk'' Poisson bracket, so that the Jacobi identity is satisfied.
\item
Given a manifest formulation of this new Poisson bracket. 
\item
Geometrized the Poisson bracket to an abstract world volume manifold.
\item
Reviewed an embedded framework to treat Dirac delta distributions 
\mb{\delta_{\Sigma}(x,\yyy)} in the presence of a boundary \mb{\pa\Sigma}.
\item
Introduced an annihilation principle and a floating Poisson bracket.
\item 
Given an action principle for Hamiltonian systems with a spatial boundaries.
\item
Discussed the relation between the boundary terms in the 
Poisson bracket and the choice of time order in a heuristic manner.
\end{itemize}


\vs
\noi
{\bf Acknowledgements}.
I would like to thank R.~Jackiw, K.~Johnson, J.~Cruz, J.~Pachos, 
M.I.~Park, R.~Schiappa and J.~Song for discussions. The research is supported
by the Danish Natural Science Research Council(SNF), grant no.~9602107.


\begin{appendix}

\setcounter{equation}{0}
\section{Various Identities}
\label{varid}

First of all, let us prove that the Heisenberg algebra reproduces
the algebra \eq{combdisentang} of partial and spatial derivatives. 
This follows from
\bea
\sum_{k,n \geq 0}\frac{q^{k}y^{n}}{n!}  P_{A(k)}~\pa^{n}
&=& P_{A}~ e^{qY}e^{\pa y} 
~=~ P_{A}~e^{\pa y} e^{qY}e^{[qY,\pa y]}
~=~ e^{\pa y} P_{A}(q) e^{q y} \cr
&=&   \sum_{m\geq 0} \sum_{k,n \geq m}\frac{(q y)^{m}}{m!} 
\frac{(\pa y)^{n-m}}{(n-m)!}  P_{A(k-m)}q^{k-m}  \cr
&=& \sum_{k,n \geq 0 } \frac{q^{k}y^{n}}{n!} \sum_{m=0}^{{\rm min}(k,n)}
\twobyone{n}{m} \pa^{n-m}P_{A(k-m)}~.
\label{reallyheis}
\eea
Next, let us check the identity
\eq{arewerollingsomemore}. The proof goes as follows:
\beq
\begin{array}{rcccl}
\pa^{k} \left[ E_{A(k)}f~ \delta \phi^{A} \right]
&=&\left. \exp\left[\pa \papa{q}\right]  
\left[ E_{A}(q)f~ \delta \phi^{A} \right]\right|_{q=0} 
&=& \int d^{d}y~  e^{\pa y} \left[ E_{A}(y)f~ \delta \phi^{A} \right] \cr
&=& \int d^{d}y~  e^{\pa y}  E_{A}(y)f~  e^{\pa y}\delta \phi^{A} 
&=& \int d^{d}y~ P_{A}(y)f~  e^{\pa y}\delta \phi^{A} \cr
&=& \left. P_{A}(q)f~\exp\left[\pa \papara{q}\right] 
\delta \phi^{A}\right|_{q=0}  
&=&  P_{A(k)}f~\pa^{k} \delta \phi^{A}~.
\end{array}
\label{wearerolling}
\eeq
Let us note the following consequences of the Heisenberg algebra
\bea
 e^{\pa y} P_{A}(y_{A})&=&P_{A}(y_{A}+y)~ e^{\pa y}~,\cr
 E_{A}(y_{A})&=&E_{A}(y_{A}+y)~ e^{\pa y}~,\cr
 E_{A}(y_{A}) E_{B}(y_{B})&=&E_{A}(y_{A}+y_{B}) P_{B}(y_{B})
~=~ P_{B}(-y_{A}) E_{A}(y_{A}+y_{B})~.
\eea 
Similary, we mention
\beq
  E_{A}(y_{A}) e^{\pa y}~=~ E_{A}(y_{A}-y)~,~~~~~~{\rm or~~equivalently}~~~~
E_{A(k)} \pa^{n}~=~E_{A(k-n)}~.
\eeq
It is worth pointing out the case \mb{k\!=\!0,n \!\neq\! 0}, which in words 
says that the  usual \mb{k\!=\!0} Euler-Lagrange derivative of a total 
derivative term is identically zero. This is hardly surprising.
Also note that the usual \mb{k\!=\!0} Euler-Lagrange operator in this 
language reads
\beq
   E_{A(k=0)}f~=~E_{A}(q\!=\!0)f~=~ \int  d^{d}y~E_{A}(y)f~.
\eeq

\setcounter{equation}{0}
\section{Proof of the Jacobi Identity (Commutative Case)}
\label{comji}

\noi
With the above machinary working, we can give a hopefully readable proof 
of the Jacobi identity. Consider three functions $f$, $g$ and $h$. Using the
fact that the usual  Euler-Lagrange derivative cannot ``feel'' a total 
derivative term, we have 
\bea
 \{ f, \{g, h \} \}&=& P_{A(a)}f ~\omega^{AB}~\pa^{a}E_{B(0)}\{g, h \}_{(0)} 
+\pa^{b} E_{A(0)}f ~\omega^{AB}~ P_{B(b)}\left[
 \pa^{d} E_{C(0)}g ~\omega^{CD}~ P_{D(d)}h \right. \cr
&&+ \left.  P_{C(c)}g ~\omega^{CD}~\pa^{c} E_{D(0)}h -\{g, h\}_{(0)} \right] 
-E_{A(0)}f~\omega^{AB}~E_{B(0)} \{g, h \}_{(0)} \cr
&=& T_{1}(f,g,h)+T_{2}(f,g,h)+T_{3}(f,g,h)-T_{4}(f,g,h)-T_{5}(f,g,h)
- (g \leftrightarrow h )~, \eea
where we have introduced a shorthand notation for the following five terms
\bea
T_{1}(f,g,h)&\equiv& P_{A(a)}f ~\omega^{AB}~ \pa^{a} (-\pa)^b \left[
 P_{B(b)}E_{C(0)}g ~\omega^{CD}~E_{D(0)}h \right]~,\cr
T_{2}(f,g,h)&\equiv& \pa^{b} E_{A(0)}f ~\omega^{AB}~ 
  P_{B(b)} P_{C(c)}g  ~\omega^{CD}~\pa^{c} E_{D(0)}h~,\cr
T_{3}(f,g,h)&\equiv& \pa^{b} E_{A(0)}f ~\omega^{AB}~ 
 P_{B(b)}\pa^{d}E_{C(0)}g ~\omega^{CD}~P_{D(d)} ~,\cr
T_{4}(f,g,h)&\equiv& \pa^{b} E_{A(0)}f ~\omega^{AB}~ 
 P_{B(b)}~E_{C(0)}g ~\omega^{CD}~E_{D(0)}\cr
T_{5}(f,g,h)&\equiv& E_{A(0)}f ~\omega^{AB}~ (-\pa)^b \left[
 P_{B(b)}E_{C(0)}g ~\omega^{CD}~E_{D(0)}h \right]~.
\eea
The Jacobi identity, containing $30$ \mb{T_{i}}-terms, now follows from 
the fact that
\beq
  T_{2}(f,g,h)~=~ T_{2}(h,g,f)~,~~~T_{1}(f,g,h)~=~T_{3}(h,g,f)~,~~~
T_{4}(f,g,h)~=~T_{5}(h,g,f)~.
\eeq
The first equation is trivial and the next two equations follows by 
rewriting in terms of Fourier transforms
\bea
T_{1}(f,g,h) &=& P_{A}(q_{A})f ~\omega^{AB}~ 
\exp\left[\pa (\papara{q_{A}}-\papa{q_{B}}) \right]
 \left. \left[ P_{B}(q_{B}) E_{C}(q_{C})g ~\omega^{CD}~
E_{D}(q_{D})h \right]\right|_{q=0} \cr
&=&  \! \! \! \!  \int d^{4d}y ~P_{A}(y_{A})f ~\omega^{AB}~ 
e^{\pa(y_{A}-y_{B})} \left[ P_{B}(y_{B}) E_{C}(y_{C})g ~\omega^{CD}~
E_{D}(y_{D})h \right] \cr
&=&  \! \! \! \! \int d^{4d}y ~P_{A}(y_{A})f ~\omega^{AB}~ 
e^{\pa(y_{A}-y_{B})} \left[ e^{-\pa y_{C}} P_{B}(y_{B}\!+\!y_{C}) 
P_{C}(y_{C})g ~\omega^{CD}~E_{D}(y_{D})h \right] \cr
&=&  \! \! \! \! \int d^{4d}y ~P_{A}(y_{A})f ~\omega^{AB}~ 
e^{\pa(y_{A}-y_{B})} P_{B}(y_{B}) P_{C}(y_{C})g ~\omega^{CD}~
e^{\pa(y_{A}+y_{C}-y_{B})} E_{D}(y_{D})h~, \cr
T_{3}(f,g,h) &=&  \left. \exp\left[\pa \papa{q_{B}}\right]  E_{A}(q_{A}) f 
~\omega^{AB}~  P_{B}(q_{B})~ \exp\left[\pa \papa{q_{D}}\right]  
E_{C}(q_{C})g ~\omega^{CD}~ P_{D}(q_{D})h\right|_{q=0} \cr
&=&  \! \! \! \! \int d^{4d}y~e^{\pa y_{B}} E_{A}(y_{A}) f 
~\omega^{AB}~  P_{B}(y_{B})~e^{\pa y_{D}}E_{C}(y_{C})g ~\omega^{CD}~
P_{D}(y_{D})h \cr
&=&  \! \! \! \! \int d^{4d}y~e^{\pa y_{B}} E_{A}(y_{A}) f 
~\omega^{AB}~ e^{\pa(y_{D}-y_{C})} P_{B}(y_{B}\!+\!y_{C}\!-\!y_{D})
P_{C}(y_{C})g ~\omega^{CD}~P_{D}(y_{D})h \cr
&=&  \! \! \! \!\int d^{4d}y~e^{\pa (y_{B}+y_{D}-y_{C})} E_{A}(y_{A})f
~\omega^{AB}~ e^{\pa(y_{D}-y_{C})} P_{B}(y_{B})P_{C}(y_{C})g 
~\omega^{CD}~P_{D}(y_{D})h~, \cr
T_{4}(f,g,h) &=&  \left. \exp\left[\pa \papa{q_{B}}\right] 
 E_{A}(q_{A})f ~\omega^{AB}~ P_{B}(q_{B})~E_{C}(q_{C})g ~\omega^{CD}~
E_{D}(q_{D})h\right|_{q=0} \cr
 &=&   \! \! \! \!  \int d^{4d}y~ e^{\pa y_{B}}  E_{A}(y_{A})f 
~\omega^{AB}~ P_{B}(y_{B})~E_{C}(y_{C})g ~\omega^{CD}~E_{D}(y_{D})h \cr
 &=&   \! \! \! \!  \int d^{4d}y~ e^{\pa y_{B}}  E_{A}(y_{A})f 
~\omega^{AB}~e^{-\pa y_{C}}P_{B}(y_{B}+y_{C})~P_{C}(y_{C})g ~\omega^{CD}~
E_{D}(y_{D})h \cr
 &=&   \! \! \! \!  \int d^{4d}y~ e^{\pa (y_{B}-y_{C})} E_{A}(y_{A})f 
~\omega^{AB}~e^{-\pa y_{C}}P_{B}(y_{B})~P_{C}(y_{C})g ~\omega^{CD}~
E_{D}(y_{D})h~, \cr
T_{5}(f,g,h)&=&   E_{A}(q_{A})f ~\omega^{AB}~  \left. 
\exp\left[-\pa \papa{q_{B}}\right] \left[ P_{B}(q_{B})E_{C}(q_{C})g 
~\omega^{CD}~E_{D}(q_{D})h \right] \right|_{q=0} \cr
&=&   \! \! \! \!  \int d^{4d}y~E_{A}(q_{A})f ~\omega^{AB}~e^{-\pa y_{B}}
 \left[ P_{B}(y_{B})E_{C}(y_{C})g ~\omega^{CD}~E_{D}(y_{D})h \right] \cr
&=&   \! \! \! \!  \int d^{4d}y~E_{A}(q_{A})f ~\omega^{AB}~e^{-\pa y_{B}}
 \left[e^{-\pa y_{C}} P_{B}(y_{B}+y_{C})P_{C}(y_{C})g ~\omega^{CD}~
E_{D}(y_{D})h \right] \cr
&=&   \! \! \! \!  \int d^{4d}y~E_{A}(q_{A})f ~\omega^{AB}~e^{-\pa y_{B}}
P_{B}(y_{B})P_{C}(y_{C})g ~\omega^{CD}~e^{\pa(y_{C}- y_{B})}E_{D}(y_{D})h~.
\eea
We have used the following shorthand notation for the integration measure
\beq
 d^{4d}y~\equiv~d^{d}y_{A}~ d^{d}y_{B}~ d^{d}y_{C}~ d^{d}y_{D}~,
\eeq
and we have performed the following change of integration variables
\beq
\fourbyone{y'_{A}}{y'_{B}}{y'_{C}}{y'_{D}}~=~
\left( \begin{array}{ccccccc}
{1}&&{0}&&{0}&&{0}\cr{0}&&{1}&&{1}&&{*}\cr
{0}&&{0}&&{1}&&{0}\cr{0}&&{0}&&{0}&&{1} \end{array} \right)
\fourbyone{y_{A}}{y_{B}}{y_{C}}{y_{D}}~,
\eeq
which has Jacobian equal to $1$. 

\proofbox

\setcounter{equation}{0}
\section{Calculus of Words}

\label{appcalcorderedtuples}

\vs
To be self-contained, we will here give a short treatment of the calculus 
with ordered index-structure, merely giving the main definitions and
formulas.

\subsection{Words}
\label{secwordss}
\noi
An ordered tuple \mb{K} (or a {\em positive word}) takes the form
\beq
K~=~(k_{1},\ldots, k_{|K|}) \in \{1, \ldots, d \}^{|K|}~.
\eeq
Here we have employed a $d$-letter {\em alphabet} \mb{\{1, \ldots, d \}}
and  \mb{|K| \in \N_{0}} denotes the {\em length} of \mb{K}.
The {\em transposed} word is \mb{K^{t} = (k_{|K|},\ldots, k_{1})}. 
We define the (non-commutative, associative) sum
of two tuples \mb{K=(k_{1},\ldots, k_{|K|})} and  
\mb{L=(\ell_{1},\ldots, \ell_{|L|})} as the concatenation
\beq
   K+L~\equiv~(k_{1},\ldots, k_{|K|},\ell_{1},\ldots, \ell_{|L|})~.
\eeq 
Obviously, the empty index set \mb{\emptyset} is the neutral element.
Formally, we can define negative words \mb{-K=-(k_{|K|})-\ldots-( k_{1})}. 
Moreover \mb{-(K+L)=-L-K}. We define a left subtraction
\mb{-K+L} as the unique solution to \mb{K+(-K+L)=L}. If the left
subtraction \mb{-K+L} is a positive word, we say that \mb{K\leq L}.
Clearly, \mb{(K+L)^{t} = L^{t} + K^{t}}.

\vs
\noi
There exists another partial order between two arbitrary positive words
\mb{K} and \mb{L}. Namely, define that $K$ preceed (or is equal to) $L$, 
written \mb{K\preceq L}, if we can obtain $K$ from $L$ by deleting some 
(possible no or all) elements in $L$. Said in a mathematical precise manner, 
there exists a (strongly) increasing index function 
\mb{\pi:\{1,\ldots, |K|\} \to \{1,\ldots, |L|\}}, called an
(orderpreserving) embedding,
such that
\bea
k_{1}&=&\ell_{\pi(1)}~,~~~~~~ \ldots~,~~~~~~  
k_{|K|}~=~\ell_{\pi(|K|)}~,\cr
   \pi(1)& <&\ldots ~<~ \pi(|K|)~.
\eea
In the affirmative case, we define the subtraction \mb{L\div K} as the tuple
of deleted entries. More precisely, in this case there exists a unique 
(strongly) increasing index function, called the complementary embedding, 
\beq
 \pi^{c} :  \{1,\ldots, |L|\!-\!|K|\} \to \{1,\ldots, |L|\}~,
\eeq
such that the images of \mb{\pi} and \mb{\pi^{c}} are disjoint, \ie
\bea
\pi\left(\{1,\ldots, |K|\}\right)&\cap&
\pi^{c}\left(\{1,\ldots, |L|\!-\!|K|\} \right) ~=~\emptyset~,  \cr
\pi^{c}(1) &<&\ldots ~<~ \pi^{c}(|L|\!-\!|K|)~. 
\eea
Then the subtraction \mb{L\div K} is defined as
\beq
   L\div K~=~(\ell_{\pi^{c}(1)}, \ldots,\ell_{\pi^{c}(|L|-|K|)} )~.
\eeq
We stress that the embedding \mb{\pi} is not necessary unique.
Therefore \mb{ L\div K} depends on the embedding  \mb{\pi}. 
For instance, in the entanglement formula \eq{covcombdisentang}, 
one should sum over all possible embeddings.
A closed expression for the degeneracy \mb{d( K \! \preceq\! L)} of 
imbeddings \mb{\pi} is not known to the author. 
By definition \mb{d( \emptyset\! \preceq\! L)=1}.
Note that \mb{K \stackrel{\pi}{\preceq} L} implies 
\mb{L\div K \stackrel{\pi^{c}}{\preceq} L}. 
In the affirmative case the pair $K$ and  \mb{L\div K} is called
\cite{stasheff} an 
{\em unshuffle} of $L$. There exist \mb{2^{|L|}} unshuffles
of $L$. Furthermore, 
\bea
(L\div K)^{t} &=& L^{t} \div K^{t}~, \cr
K~ \preceq~ L~  \preceq~ N   ~~~&\Rightarrow&~~~~   
N \div L~=~ ( N \div K) \div (L \div K)~.
\eea
A {\em shuffle} \mb{K \# M} of two positive words \mb{K} and \mb{M} 
is defined as the opposite of an unshuffle in the sense that it
is a solution \mb{X} to \mb{X \div K = M}. Clearly, the number of shuffles 
for fixed \mb{K} and \mb{M} is \mb{\twobyone{|K|+|M|}{|M|}}. 
The number of solutions \mb{X} to  \mb{L \div X = M} for fixed positive 
\mb{M}, \mb{L} with \mb{M \preceq L} is  \mb{d( M \! \preceq\! L)} .

\subsection{Alphabets of Operators}

\noi
A $d$-letter alphabet of operators (or more generally, of associative 
abstract algebra elements), is just $d$ operators
\mb{A=(A_{1}, \ldots, A_{d})}. The sum and the product (\ie usually the 
operator composition) of two alphabets are defined letterwise
\beq
A+B~=~(A_{1}+B_{1}, \ldots, A_{d}+B_{d})~,~~~~~~~~~ 
A B~=~(A_{1} B_{1}, \ldots, A_{d}B_{d})~,
\label{alphasumprod}
\eeq 
respectively.

\subsection{Words of Operators}

\noi
If we have an alphabet of operators \mb{A=(A_{1}, \ldots, A_{d})}
we can form words of operators
\beq
A^{K}~=~A_{k_{1}} \ldots A_{k_{|K|}}~. 
\label{defword}
\eeq
We invoke the convention that \mb{A_{K}=A^{(K^{t})}} denotes the transposed 
word. The empty word operator \mb{A^{\emptyset}=1} is the identity.
Operators for non-positive words, \ie for words containing negative letters, 
are declared to be zero. Concatenation leads to a non-commutative product 
\beq
A^{N}  A^{M}~=~ A^{N+M}
\label{concaprod}
\eeq
 between words 
(and \mb{A_{N}   A_{M}= A_{M+N}} for the transposed). It coincides 
with the letterwise multiplication. But this is not the only associative 
product of words. Shuffling leads to a commutative \mb{*}-product
\beq
\frac{A^{K}}{|K|!} *  \frac{A^{L}}{|L|!}
~=~\sum_{N=~K \# L} \frac{A^{N}}{|N|!}~, ~~~~~~~~~~~
\left(~~ A^{K} * 1~=~A^{K}~~ \right)~,
\label{starprod}
\eeq
where the sum is over possible shuffles \mb{K \# L}. The definition of
the \mb{*}-product is extended by \mb{\C}-bilinarity. The concatenation 
product and the \mb{*}-product coincide for commutative alphabets.
The following binomial relation is a consequence of the special features of 
unshuffles, cf.\ \eq{alphasumprod} and \eq{defword},
\beq
[A_{i},B_{j}]~=~0~~~\Rightarrow~~~~ 
(A+B)^{N} ~=~   \sum_{M=\emptyset}^{M \preceq N} A^{M} B^{N \div M}~.
\label{binomialformula}
\eeq
We can also define a non-commutative, associative sum by the following 
binomial expression   
\beq
\frac{(A \# B)^{N}}{|N|!}  ~=~  \sum_{K,L \succeq \emptyset}^{K+L=N}
~ \frac{A^{K}}{|K|!}~\frac{B^{L}}{|L|!} ~.
\label{binomialformula2}
\eeq 
This sum  ``\mb{\#}'' and  the usual sum ``\mb{+}'' do {\em not} coincide 
in general, not even for commutative alphabets. But see the equation 
\eq{hegnusual} below for further relationship.

\vs
\noi
Note that the definition \eq{starprod} makes no use of the algebra 
multiplication. It only needs a vector space with a basis of vectors 
\mb{A^{K}} indexed by words \mb{K}. The definition of the \mb{\#}-sum 
\eq{binomialformula2} needs an algebra of words \mb{A^{K}}, but it is 
irrelevant whether \mb{A^{K}} is a composite object of more elementary
letters or not.  The same remark could be made about formulas 
\eq{concaprod} and \eq{binomialformula}, that in the minimalistic 
interpretation becomes definitions. Even in the case where there exists a
letterwise algebra multiplication, we will often use the binomial 
formula \eq{binomialformula}, also known as a convolution, 
which makes sense even for mutually non-commuting alphabets.

\vs
\noi
We have that \mb{-(A \# B)=(-A) \# (-B) } and
\beq 
 \left.\begin{array}{rcl} 
 (A^{N})^{t}  &=&  A^{(N^{t})}  \cr (B^{N})^{t}  &=&  B^{(N^{t})} 
\end{array} \right\} ~~~~\Rightarrow~~~~~
  [(A \# B)^{N}]^{t}= (B \# A)^{(N^{t})}~.
\eeq
Sometimes we will also need to define 
\mb{\frac{(A^{t} \# B)^{N}} { |N|!} \equiv
\sum_{K,L \succeq \emptyset}^{K+L=N}
 \frac{A^{(K^{t})}} { |K|!} ~ 
\frac{ B^{L} }{ |L|!}}, etc.

\subsection{Functions of Operators}

\noi
For an analytic function \mb{f(x)= \sum_{n=0}^{\infty} a_{n} x^{n}}, we 
define \mb{f(A)= \sum_{N=\emptyset}^{\infty} a_{|N|} A^{N}}. 
In particular, the exponential of an alphabet is
\beq
   \exp(A)~=~ \sum_{N=\emptyset}^{\infty} \frac{A^{N}}{|N|!}~.
\eeq
We have that 
\bea
[A_{i},A_{j}]~=~0~~~&\Rightarrow&~~~~ 
\exp(A)~=~\exp(A_{1})\ldots \exp(A_{d})~, \cr
[A_{i},B_{j}]~=~0~~~&\Rightarrow&~~~~ \exp(A+B)~=~\exp(A)\exp(B)~.
\eea
Also we have the important orthogonality relation
\beq
((-A)^{t}+A)^{N}~\equiv~
\sum_{M=\emptyset}^{M \preceq N} (-A)^{(M^{t})}~A^{N\div M} 
~=~  \delta_{N,\emptyset}~\equiv~0^{N}~.   
\eeq
This property leads to the vital relation \mb{e^{A} e^{-A}=1} 
even for a non-commutative alphabet. Similary, we have 
\mb{((-A)^{t} \# A)^{N}=0^{N}}.

\vs
\noi
In practice, we only use \mb{f(A B)} for a product of two 
alphabets. (Dummy indices usually come in pairs.) It is convenient to define 
a ``dot product'' notation
\beq
  f(A \ccdot B)~=~ \sum_{N=\emptyset}^{\infty} a_{|N|} A^{N} B^{(N^{t})}~,
\eeq
implementing a transposition of one of the alphabets. 
Also \mb{ f((-A)^{t} \ccdot B)\equiv f((-A) B)}.
We have the following inversion relation for the concatenation product
\beq
[A_{i},B_{j}]~=~0~~~~\Rightarrow~~~~~ 
\exp\left[(-A) \ccdot B \right]   \exp(A \ccdot B)  ~=~1~.
\eeq
This should be compared with the inversion relation for the the mixed
case of an implicitly written concatenation product and a \mb{*}-product
\beq
[A_{i},B_{j}]~=~0~~~~\Rightarrow~~~~~ 
\exp\left[(-A)^{t} \ccdot B \right]  ~ *_{B}~    \exp(A \ccdot B) ~=~1~.
\eeq 
We have the following distributive laws
\beq
\left.\begin{array}{rcl} 
[A_{i},B_{j}]&=&0 \cr 
[A_{i},C_{j}]&=&0 \cr 
[B_{i},C_{j}]&=&0
\end{array} \right\} ~~~~\Rightarrow~~~~~
\left\{\begin{array}{rcl}
\exp\left[(-A)^{t} \ccdot (B \# C)\right]
&=& \exp((-A)^{t} \ccdot B) ~ \exp((-A)^{t} \ccdot C) \cr 
\exp\left[A \ccdot (B \# C)\right]
&=& \exp(A \ccdot C) ~ \exp( A \ccdot B)  \cr 
\exp\left[(A + B) \ccdot C \right]
&=& \exp(A \ccdot C)~ *_{C}~ \exp(B\ccdot  C) \cr 
&=& \exp(B\ccdot  C)~ *_{C}~ \exp(A\ccdot  C)
\end{array} \right.~.
\eeq
Note the reversed order in the second equation.
The sum  ``\mb{\#}'' and  the usual sum ``\mb{+}'' coincide loosely speaking
in average. More precisely, 
\beq
[A_{i},A_{j}]~=~0~~~~\Rightarrow~~~~~ 
\exp\left[A \ccdot( B \# C) \right] ~=~  \exp\left[A \ccdot (B+C)\right]~.
\label{hegnusual}
\eeq

\subsection{Trace and Fourier Analysis}

We define a trace on the vector space of words, which is constructed from two 
mutually commuting freely generated alphabets \mb{A} and \mb{B}:
\beq
\int d^{d}A~d^{d}B~e^{(-A)^{t}\ccdot B}~ *_{B}~  A^{N}~ B_{M}~\equiv~ 
{\rm Tr}(A^{N}~ B_{M})~=~ |N|!~\delta^{N}_{M}~,
\eeq
and extend by \mb{\C}-bilinarity.
As the first equality suggests, we will sometimes use a suggestive 
notation for the trace borrowed from the Fourier analysis in the usual 
commutative case. One can take this analogy quite far. We do not 
give any meaning to the position of the measure \mb{d^{d}A~d^{d}B}, \ie
it is taken to commute with everything.
A theoretically perhaps more convenient form is
\beq
\int d^{d}A~d^{d}B~\exp \left[(-A-A')^{t} \ccdot (B\#B') \right]~=~1~.
\eeq
{}From here it follows trivially that the integration measure 
\mb{d^{d}A~d^{d}B} is translation invariant under \mb{ A \to A+ A'},
\mb{ B \to B \# B'}  (but not under \mb{B \to B' \# B}!).

\subsection{Differentiation of Words}
\label{nonass}

\noi
Consider the differential alphabet 
\mb{\papa{A}= (\papa{A_{1}}, \ldots, \papa{A_{d}})} of
freely generated associative algebra elements \mb{A_{i}}.
Let us define differentiation  at the level of words ( \ie not letterwise), 
as 
\beq
\frac{1}{|K|!}  \papa{A^{K}} \left[ A^{L} \right] 
~=~\sum_{ \pi: K \preceq L}  A^{L \stackrel{\pi}{\div} K}~,~~~~~~~~~~~
\left(~~\papa{A^{\emptyset}}\left[ 1 \right] ~=~1~~ \right)~,
\label{diffda} 
\eeq
where the sum is over possible embeddings \mb{\pi}. Extend the definition
by \mb{\C}-bilinarity. One of the main motivations behind this definition 
is to implement the Taylor formula 
\bea
[A_{i},B_{j}]~=~0~~~~~~~&\Rightarrow&~~~~ 
 \exp \left[B \ccdot  \papa{A}\right]  f(A)~=~f(B+A) ~.
\eea
The composition of derivatives is described by the \mb{*}-product
\beq
 \frac{1}{|K|!}  \papa{A^{K}}~  \frac{1}{|L|!}  \papa{A^{L}} 
~=~ \sum_{N=~K \# L} \frac{1}{|N|!}  \papa{A^{N}}
~\equiv~ \frac{1}{|K|!}  \papa{A^{K}}~* ~ \frac{1}{|L|!}  \papa{A^{L}} ~.
\eeq
As consequences, the derivatives are associative and commutative
wrt.\ composition. They enjoy the following properties 
\bea
[A_{i},B_{j}]~=~0~~~~~~~~&\Rightarrow&~~~~ 
\exp\left[(-B)^{t} \ccdot \papa{A} \right] 
\exp\left[B \ccdot \papa{A}\right] ~=~1~, \cr\cr
\left.\begin{array}{rcl} 
[A_{i},C_{j}]&=&0 \cr   
[B_{i},C_{j}]&=&0
\end{array} \right\} ~~~&\Rightarrow&~~~~ 
\exp\left[(B+C) \ccdot \papa{A} \right] ~=~
 \exp \left[B \ccdot \papa{A} \right]
\exp \left[C \ccdot \papa{A} \right]~.
\eea
We can of course implement the concatenation product for the
differential. However, in practical calculations it plays no role.
The derivatives \eq{diffda} do not satisfy the Leibnitz' rule. 

\vs
\noi
The are other kinds of differential alphabets. In our case, we have the 
covariant derivatives \mb{D_{i}} that act on words \mb{(-Y)_{K}} 
according to
\beq
   [D_{i}, (-Y)_{K} ]
 ~=~\left\{ \begin{array}{l}
    |K|~ (-Y)_{-(i) + K}    ~~{\rm if}~~ (i) \leq  K~, \cr
0 ~~{\rm otherwise}~.
\end{array} \right.           
\eeq
(One can consider the words \mb{(-Y)_{K}} as originating from an
associative algebra alphabet \mb{(-Y^{1}, \ldots ,-Y^{d})} that behaves 
non-associatively wrt.\ the \mb{D_{i}}'s, but it is unnecessary.)
The covariant derivatives satisfies Leibnitz' rule on functions:
\beq
   D^{K}(fg)~=~\sum_{M=\emptyset}^{M \preceq K}  D^{M}f~ D^{K\div M}g~.
\eeq
This can be recasted into the Taylor-like form
\mb{e^{D \ccdot y}(fg)~=~e^{D \ccdot y}f~*_{y}~e^{D \ccdot y}g}.

\setcounter{equation}{0}
\section{Proof of the Jacobi Identity (Non-Commutative Case)}
\label{noncomji}

\noi
We have 
\bea
 \{ f, \{g, h \} \}&=& P_{A(A)}f ~\omega^{AB}~D^{A} E_{B(0)} \{g, h \}_{(0)} 
+D^{B} E_{A(0)}f ~\omega^{AB}~ P_{B(B)}\left[
 D^{D} E_{C(0)}g ~\omega^{CD}~ P_{D(D)}h \right. \cr
&&+ \left.  P_{C(C)}g ~\omega^{CD}~D^{C} E_{D(0)}h -\{g, h \}_{(0)} \right] 
-E_{A(0)}f~\omega^{AB}~E_{B(0)} \{g, h \}_{(0)} \cr
&=& T_{1}(f,g,h)+T_{2}(f,g,h)+T_{3}(f,g,h)-T_{4}(f,g,h)-T_{5}(f,g,h)
- (g \leftrightarrow h )~, \eea
where we have introduced a shorthand notation for the following five terms
\bea
T_{1}(f,g,h)&\equiv& P_{A(A)}f ~\omega^{AB}~ D^{A} (-D)^{B^{t}} \left[
 P_{B(B)}E_{C(0)}g ~\omega^{CD}~E_{D(0)}h \right]~,\cr
T_{2}(f,g,h)&\equiv& D^{B} E_{A(0)}f ~\omega^{AB}~ 
  P_{B(B)} P_{C(C)}g  ~\omega^{CD}~D^{C} E_{D(0)}h~,\cr
T_{3}(f,g,h)&\equiv& D^{B} E_{A(0)}f ~\omega^{AB}~ 
 P_{B(B)}D^{D}E_{C(0)}g ~\omega^{CD}~P_{D(D)} ~,\cr
T_{4}(f,g,h)&\equiv& D^{B} E_{A(0)}f ~\omega^{AB}~ 
 P_{B(B)}~E_{C(0)}g ~\omega^{CD}~E_{D(0)}\cr
T_{5}(f,g,h)&\equiv& E_{A(0)}f ~\omega^{AB}~ (-D)^{B^{t}} \left[
 P_{B(B)}E_{C(0)}g ~\omega^{CD}~E_{D(0)}h \right]~.
\eea
Here we have chosen to use the same index symbol \mb{A}, \mb{B}, \mb{C} and
\mb{D} to label the indices of the fields \mb{\phi} and the words.
It should not lead to any ambiguities, and it hopefully becomes easier to 
grasp the index structure. 
The Jacobi identity, containing $30$ \mb{T_{i}}-terms, now follows from 
the fact that
\beq
  T_{2}(f,g,h)~=~ T_{2}(h,g,f)~,~~~T_{1}(f,g,h)~=~T_{3}(h,g,f)~,~~~
T_{4}(f,g,h)~=~T_{5}(h,g,f)~.
\label{boilsdownto}
\eeq
The first equation is trivial and the next two equations follows by rewriting 
in terms of Fourier transforms
\bea
T_{1}(f,g,h) &=& P_{A}\qqa f ~\omega^{AB}~ 
\exp \left[D \ccdot \papara{q_{A}} \right]
\exp \left[(-D)^{t}\ccdot \papa{q_{B}} \right]
\left. \left[ P_{B}\qqb  E_{C}\qqc g ~\omega^{CD}~
E_{D}\qqd h \right]\right|_{q=0} \cr 
&=&{\rm Tr}_{DCBA}  ~P_{A}\qqa f ~\omega^{AB}~ e^{D \ccdot y_{A}} 
e^{(-D)^{t} \ccdot y_{B}} \left[ P_{B}\qqb  e^{(-D)^{t} \ccdot y_{C}}     
P_{C}\qqc g ~\omega^{CD}~E_{D}\qqd h \right] \cr \cr
&=&{\rm Tr}_{DCBA} ~P_{A}\qqa f ~\omega^{AB}~ e^{D \ccdot y_{A}} 
e^{(-D)^{t} \ccdot y_{B}}  \left[ e^{(-D -q_{B})^{t} \ccdot y_{C}} 
P_{B}\qqb  P_{C}\qqc g   ~\omega^{CD}~E_{D}\qqd h \right] \cr
&=&{\rm Tr}_{DCB'A}  ~P_{A}\qqa f ~\omega^{AB}~ e^{D \ccdot y_{A}} 
e^{(-D)^{t} \ccdot y'_{B}} P_{B}\qqb  P_{C}\qqc g ~\omega^{CD}~
e^{D \ccdot y_{A}} e^{(-D)^{t} \ccdot  y'_{B}}  e^{D \ccdot y_{C}} 
E_{D}\qqd h~, \cr
T_{3}(f,g,h) &=&  \left.\exp\left[D \ccdot \papa{q_{B}}\right] E_{A}\qqa f 
~\omega^{AB}~  P_{B}\qqb \exp\left[D \ccdot    \papa{q_{D}}\right]  
E_{C}\qqc g ~\omega^{CD}~ P_{D}\qqd h \right|_{q=0} \cr
&=& {\rm Tr}_{DCA}~ e^{D \ccdot y_{B}} E_{A}\qqa  f~\omega^{AB}~ 
{\rm Tr}_{B}~ P_{B}\qqb e^{D \ccdot y_{D}}  e^{(-D)^{t} \ccdot y_{C}}     
P_{C}\qqc g ~\omega^{CD}~P_{D}\qqd h  \cr \cr
&=& {\rm Tr}_{DCA}~e^{D \ccdot y_{B}} E_{A}\qqa  f ~\omega^{AB}~ 
{\rm Tr}_{B} ~ e^{(D + q_{B}) \ccdot y_{D}} e^{(-D-q_{B})^{t} 
\ccdot y_{C}} P_{B}\qqb P_{C}\qqc g~\omega^{CD}~P_{D}\qqd h \cr \cr
&=& {\rm Tr}_{DCA}~e^{D \ccdot  y_{D}} e^{(-D)^{t} \ccdot y_{C}}
e^{D \ccdot y'_{B}}  E_{A}\qqa f ~\omega^{AB}~{\rm Tr}_{B'}~
e^{D \ccdot y_{D}} e^{(-D)^{t} \ccdot y_{C}} P_{B}\qqb P_{C}\qqc g 
~\omega^{CD}~P_{D}\qqd h  ~, \cr
T_{4}(f,g,h) &=&  \left. \exp\left[D \ccdot  \papa{q_{B}}\right] 
E_{A}\qqa f ~\omega^{AB}~ P_{B}\qqb E_{C}\qqc g ~\omega^{CD}~
E_{D}\qqd h\right|_{q=0} \cr
&=&{\rm Tr}_{DCA}~e^{D \ccdot y_{B}} E_{A}\qqa f ~\omega^{AB}~{\rm Tr}_{B}~ 
P_{B}\qqb e^{(-D)^{t}\ccdot y_{C}} P_{C}\qqc g~\omega^{CD}~E_{D}\qqd h\cr\cr
&=&{\rm Tr}_{DCA} e^{D \ccdot y_{B}} E_{A}\qqa f ~\omega^{AB}~{\rm Tr}_{B}~
e^{(-D-q_{B})^{t} \ccdot y_{C}} P_{B}\qqb P_{C}\qqc g ~\omega^{CD}~
E_{D}\qqd h  \cr \cr
&=&{\rm Tr}_{DCA}~e^{(-D)^{t} \ccdot y_{C}}  e^{D \ccdot y'_{B}}E_{A}\qqa f 
~\omega^{AB}~{\rm Tr}_{B'}~e^{(-D)^{t} \ccdot y_{C}}P_{B}\qqb P_{C}\qqc g 
~\omega^{CD}~E_{D}\qqd h~, \cr
T_{5}(f,g,h)&=&   E_{A}\qqa f ~\omega^{AB}~  \left. 
\exp\left[(-D)^{t} \ccdot  \papa{q_{B}}\right] 
\left[ P_{B}\qqb E_{C}\qqc g 
~\omega^{CD}~E_{D}\qqd h \right] \right|_{q=0} \cr
&=&{\rm Tr}_{DCBA} ~E_{A}\qqa f ~\omega^{AB}~e^{(-D)^{t} \ccdot y_{B}}
 \left[ P_{B}\qqb  e^{(-D)^{t} \ccdot y_{C}}  P_{C}\qqc g 
~\omega^{CD}~E_{D}\qqd h \right] \cr \cr
&=& {\rm Tr}_{DCBA}~E_{A}\qqa f ~\omega^{AB}~e^{(-D)^{t} \ccdot y_{B}}
 \left[e^{(-D-q_{B})^{t} \ccdot y_{C}} P_{B}\qqb P_{C}\qqc g ~\omega^{CD}~
E_{D}\qqd h \right] \cr \cr
&=&{\rm Tr}_{DCB'A} ~E_{A}\qqa f ~\omega^{AB}~e^{(-D)^{t} \ccdot y'_{B}}
P_{B}\qqb P_{C}\qqc g ~\omega^{CD}~ e^{(-D)^{t} \ccdot y'_{B}}
e^{D \ccdot y_{C}}   E_{D}\qqd h~.
\eea
We have suppressed \mb{*}-products among the \mb{y}-variables, and shorten
\mb{P_{A}=P_{A}(q_{A}) }, \mb{P_{B}=P_{B}(q_{B}) }, \mb{\ldots},  
\mb{E_{A}=E_{A}(q_{A}) }, etc.
The trace can be written more suggestively as  
\beq
{\rm Tr}_{DCBA}~=~~\int d^{4d}q~d^{4d}y~ e^{(-q_{D})^{t}\ccdot y_{D}} 
e^{(-q_{C})^{t}\ccdot y_{C}}  e^{(-q_{B})^{t}\ccdot y_{B}} 
e^{(-q_{A})^{t}\ccdot y_{A}}~.
\eeq
We have performed the following type of translation of the integration 
variables
\beq
y'_{B}~=~y_{B} \# y_{C}~~~~~{\rm or}~~~~~~~
y'_{B}~=~y_{B} \# (-y_{D})^{t} \# y_{C}~.
\eeq
Note that after the shift of integration variables the 
\mb{y}-alphabets do no longer mutually commute. However, one may 
convince oneself that the integrations can be unwind, and
we can consistently declare them to mutually commute also 
in the new variables. Finally to prove \eq{boilsdownto}, one should
relabel dummy variables \mb{ABCD \to DCBA}.

\proofbox

\setcounter{equation}{0}
\section{Proof of the Jacobi Identity (Floating Type)}
\label{singji}

\noi
We now turn to the proof of the Jacobi identity for the floating
Poisson bracket, cf.\ Eq.\ \eq{fullpb4}.
Consider the local functionals of Subsection~\ref{seclft}. We assume that
\mb{D_{i}\rho=0}.
Suppressing the integrations, we have 
\bea
 \{ f, \{g, h \} \}&=& 
\frac{E_{A(0)}(\chii_{\epsilon}f)}{\chii_{\epsilon}}~\omega^{AB}~
E_{B(0)}\left[  E_{C(0)}(\chii_{\epsilon}g)~\omega^{CD}~
\frac{E_{D(0)}(\chii_{\epsilon}h)}{\chii_{\epsilon}} \right] \cr 
 &=& T(f,g,h) ~-~ (g \leftrightarrow h )~, 
\eea
where 
\bea
T(f,g,h)&\equiv& 
\frac{E_{A(0)}(\chii_{\epsilon}f)}{ \chii_{\epsilon} }~\omega^{AB}~
(-D)^{B^{t}} \left[ P_{B(B)} E_{C(0)}(\chii_{\epsilon}g)~\omega^{CD}~
\frac{E_{D(0)}(\chii_{\epsilon}h)}{\chii_{\epsilon}} \right] \cr
&=&\frac{E_{A}(\chii_{\epsilon}f)}{ \chii_{\epsilon}}~
\omega^{AB}~\exp \left[(-D)^{t}\ccdot \papa{q_{B}} \right]
\left. \left[ P_{B}\qqb  E_{C}\qqc  (\chii_{\epsilon}g)~\omega^{CD}~
\frac{E_{D}\qqd (\chii_{\epsilon}h)}{\chii_{\epsilon}}  
\right]\right|_{q=0} \cr 
&\sim&{\rm Tr}_{DCA}~  e^{\frac{1}{2} D \ccdot y_{B}}  
\frac{E_{A}(\chii_{\epsilon}f)}{ \chii_{\epsilon}}~\omega^{AB}~
{\rm Tr}_{B}~ e^{\frac{1}{2}(-D)^{t} \ccdot y_{B}}  \left[ P_{B}\qqb  
e^{(-D)^{t} \ccdot y_{C}} P_{C}\qqc(\chii_{\epsilon}g)~\omega^{CD}~  
\frac{E_{D}\qqd (\chii_{\epsilon}h)}{\chii_{\epsilon}} \right] \cr 
&=&{\rm Tr}_{DCA}~  e^{\frac{1}{2} D \ccdot y_{B}}  
\frac{E_{A}(\chii_{\epsilon}f)}{ \chii_{\epsilon}}~\omega^{AB}~
{\rm Tr}_{B}~e^{\frac{1}{2}(-D)^{t} \ccdot y_{B}}  
\left[e^{(-D-q_{B})^{t} \ccdot y_{C}}  
P_{B}\qqb  P_{C}\qqc(\chii_{\epsilon}g)~\omega^{CD}~  
\frac{E_{D}\qqd (\chii_{\epsilon}h)}{\chii_{\epsilon}} \right] \cr
&=&{\rm Tr}_{DCA}~ e^{\frac{1}{2}(-D)^{t} \ccdot y_{C}} 
e^{\frac{1}{2}D \ccdot y'_{B}} 
\frac{E_{A}(\chii_{\epsilon}f)}{\chii_{\epsilon}}~\omega^{AB}~\cr
&&~~~~~~~~~~ \times~~~  
{\rm Tr}_{B'}~e^{\frac{1}{2}(-D)^{t} \ccdot y'_{B}} 
e^{\frac{1}{2} D \ccdot y_{C}} \left[e^{(-D)^{t} \ccdot y_{C}}  
P_{B}\qqb  P_{C}\qqc(\chii_{\epsilon}g)~\omega^{CD}~ 
\frac{E_{D}\qqd (\chii_{\epsilon}h)}{\chii_{\epsilon}} \right]\cr
&=&{\rm Tr}_{DCBA}~ e^{\frac{1}{2}(-D)^{t} \ccdot y_{C}} 
e^{\frac{1}{2} D \ccdot y_{B}} 
\frac{E_{A}(\chii_{\epsilon}f)}{ \chii_{\epsilon}}~\omega^{AB}~
e^{\frac{1}{2}(-D)^{t} \ccdot y_{B}} e^{\frac{1}{2}(-D)^{t} \ccdot y_{C}}  
P_{B}\qqb  P_{C}\qqc(\chii_{\epsilon}g)~\omega^{CD}~ \cr
&&~~~~~~~~~~ \times~~~  
e^{\frac{1}{2}(-D)^{t} \ccdot y_{B}} e^{\frac{1}{2} D \ccdot y_{C}} 
\frac{E_{D}\qqd (\chii_{\epsilon}h)}{\chii_{\epsilon}}
~~\sim~~T(h,g,f)~.
\label{tsymcalc}
\eea
The \mb{\sim} indicates that the equality holds up to total derivative terms.
They are unphysical terms living far away from the bounded physical region 
\mb{\Sigma}, and therefore vanishing. In the last step we substituted 
\mb{y'_{B}=y_{B} \# y_{C}}. The annihilation principle will not change the 
fact, that the Jacobi identity is fulfilled, because all annihilated terms 
appear in pairs with opposite sign. 

\proofbox

\noi
Let us now turn to more general functionals \mb{F(u)},  \mb{G(v)} and
\mb{H(w)}.  We have
\bea
 \{ F(u), \{G(v), H(w) \} \}&=& \int \rho(x) d^{d}x~\rho(y) d^{d}y~
F^{B}(x)~ \dedetwo{}{\phi^{B}(x)} \left[ \chii_{\epsilon}(x)~
\dedetwo{[\chii_{\epsilon}(y)G(v)]}{\phi^{C}(y)}~H^{C}(y) \right] \cr 
 &=& \int \rho(x) d^{d}x~\rho(y) d^{d}y~T(F,G,H) 
~-~ (G(v) \leftrightarrow H(w) )~. 
\eea
Here we have applied the following shorthand notation
\beq
F^{B}(x)~=~\dedetwo{[\chii_{\epsilon}(x)F(u)]}{\phi^{A}(x)}
\frac{\omega^{AB}}{\chii_{\epsilon}(x)} ~,~~~~~~~
H^{C}(y)~=~\frac{\omega^{CD}}{\chii_{\epsilon}(y)}
\dedetwo{[\chii_{\epsilon}(y)H(w)]}{\phi^{D}(y)}~,
\eeq
and
\bea
T(F,G,H)&\equiv& F^{B}(x)~  (-D_{(x)})^{B^{t}}\left[
 \chii_{\epsilon}(x)~\papatwo{}{\phi^{B(B)}(x)}
\dedetwo{[\chii_{\epsilon}(y)G(v)]}{\phi^{C}(y)} \right] H^{C}(y) \cr 
&=& F^{B}(x)~ (-D_{(x)})^{B^{t}}  \left[ \chii_{\epsilon}(x)~
\papatwo{}{\phi^{B(B)}(x)}\sum_{j=1}^{r} (-D_{(y)})^{C^{t}}
\frac{\delta(y\!-\!v_{(j)})}{\rho(y)} 
P_{C(C)}^{(v_{(j)})} G(v)  \right] H^{C}(y)  \cr 
&=& T_{1}(F,G,H)+T_{2}(F,G,H)~,
\label{t12split}
\eea
where the last equality will be explained below. We distinguish between the 
so-called {\em inner} \mb{j}-terms \mb{j=1,\ldots,s}, where the spatial
\mb{D_{(y)}}-differentiation are applied on the \mb{P_{C}}-derivatives of the 
\mb{G}-functional before the partial derivative \mb{P_{B}}, and on the other 
hand the so-called {\em outer} \mb{j}-terms, where the order is the opposite.
For each inner \mb{j}-term, we may write 
\mb{G(v)=\int \rho(v_{(j)}) d^{d}v_{(j)}~g_{j}(v)}. 
Together, the  diagonal piece of inner \mb{j}-terms becomes
\bea
T_{1}(F,G,H)&\equiv& F^{B}(x)\left[ (-D_{(x)})^{B^{t}} 
 \sum_{j=1}^{s}\frac{\delta(x\!-\!v_{(j)})}{\rho(x)}  P_{B(B)}^{(v_{(j)})}
(-D_{(v_{(j)})})^{C^{t}} 
P_{C(C)}^{(v_{(j)})} g_{j}(v)_{|_{v_{(j)}=y}}  \right] H^{C}(y) \cr
&=& F^{B}(x)\left[ (-D_{(x)})^{B^{t}} 
 \sum_{j=1}^{s}
\frac{\delta(x\!-y)}{\rho(y)} P_{B(B)}^{(v_{(j)})}
(-D_{(v_{(j)})})^{C^{t}} P_{C(C)}^{(v_{(j)})} g_{j}(v)_{|_{v_{(j)}=y}}   
\right] H^{C}(y)~, \cr &&
\eea
The \mb{y}-integration may be explicitly performed in the diagonal 
\mb{T_{1}}-term: 
\beq
\int\rho(y) d^{d}y~T_{1}(F,G,H)~=~ F^{B}(x) (-D_{(x)})^{B^{t}}\left[ 
\sum_{j=1}^{s} P_{B(B)}^{(v_{(j)})}
(-D_{(v_{(j)})})^{C^{t}}P_{C(C)}^{(v_{(j)})}
g_{j}(v)_{|_{v_{(j)}=x}} 
~H^{C}(x)\right] ~.
\eeq
It may now be treated similarly to the local case discussed in equation 
\eq{tsymcalc}. The rest of the terms appearing in \eq{t12split} 
can be organized so that they are manifestly symmetric in \mb{F} and 
\mb{H}, and hence do not effectively contribute to the Jacobi identity: 
\beq 
T_{2}(F,G,H)~\equiv~ F^{B}(x)\left[ (-D_{(x)})^{B^{t}} (-D_{(y)})^{C^{t}}
\sum_{i, j}{}^{'}\frac{\delta(x\!-\!v_{(i)})}{\rho(x)} 
\frac{\delta(y\!-\!v_{(j)})}{\rho(y)}  
P_{B(B)}^{(v_{(i)})} P_{C(C)}^{(v_{(j)})} G(v)  \right] H^{C}(y)~.
\eeq
(The prime \mb{'} indicates that the above inner diagonal \mb{j}-terms 
should not be included in the \mb{T_{2}}-sum.)

\proofbox

\section{Realization of Derivatives}

\noi
In this section we will like to incode in an alternative manner the 
information about which part of an expression that are hit by a derivative. 
Instead of the usual practice of indicating this with arbitrarily many arrows,
we define a linear chain of operators containing the same information. 
Consider some expression, where upon the derivatives \mb{D_{i}} act.
{\em Notation: For simplicity, we assume that the covariant 
derivatives commute, \ie that the curvature vanishes.} 
First of all, we assign  names \mb{\alpha} to the derivatives 
\mb{D_{i} \leadsto D_{(\alpha)i}} so that we may distinguish them. 
For all derivatives \mb{D_{(\alpha)i}}, let us 
introduce a scalar elements \mb{a_{(\alpha)}} and a vector element 
\mb{b_{(\alpha)}^{i}} that obey
\beq
a_{(\alpha)} a_{(\alpha)}~=~1~,~~~~~
b_{(\alpha)}^{i} b_{(\alpha)}^{i}~=~0~,~~~~~
b_{(\alpha)}^{i} a_{(\alpha)} ~=~-b^{i}~,~~~~~
a_{(\alpha)} b_{(\alpha)}^{i} ~=~ b_{(\alpha)}^{i}~. 
\eeq 
The notions scalar and vector refer to their properties under 
coordinate transformations. We have the following commutator relations
\beq
\left[a_{(\alpha)}, a_{(\beta)}\right]~=~0~,~~~~~
\left[b_{(\alpha)}^{i}, b_{(\beta)}^{j}\right]~=~0~,~~~~~
\left[a_{(\alpha)}, b_{(\alpha)}^{j}\right]_{+}~=~0~,~~~~~
\left[a_{(\alpha)}, b_{(\beta)}^{j}\right]
~=~2 \delta_{\alpha\beta}~b_{(\beta)}^{j}~.
\eeq
Furthermore, we introduce a Berezin-like integration \mb{\int db_{(\alpha)i}}.
In the following, \mb{f} and \mb{f'} denote functions which do not depend on 
\mb{a_{(\alpha)}} and \mb{b_{(\alpha)}^{i}}, but possibly on other 
\mb{a}'s and \mb{b}'s.  We will assign the value
\beq
\int db_{(\alpha)i}~b_{(\alpha)}^{j}~ f ~=~\delta_{i}^{j}~f~,~~~~~
\int db_{(\alpha)i}~f ~=~0~.
\eeq 
We emphasize that we have not defined the integral \mb{\int db_{(\alpha)i}}
if there are \mb{a_{(\alpha)}} elements present in the integrand. 
Let us also declare
\beq
\left[b_{(\alpha)}^{i},f\right]~=~0~,~~~~\left[f,f'\right]~=~0~,~~~
\left[f,~a_{(\alpha)}~f'~a_{(\alpha)}\right]~=0~,
\eeq
Now we are ready to define the transition from derivatives to \mb{a}'s 
and \mb{b}'s. The derivative \mb{D_{(\alpha)i}f} may now be written as
\beq
 D_{(\alpha)i}f~~\to~~
\int db_{(\alpha)i}~ a_{(\alpha)}~ f~ a_{(\alpha)}
~=~\int db_{(\alpha)i} \left[a_{(\alpha)},f\right] a_{(\alpha)}~.
\eeq
This on the other hand is evaluated by declaring
\beq
-\left[a_{(\alpha)},f \right]~ \equiv ~\left[f,a_{(\alpha)} \right]
 ~~=~~\sum_{i} b_{(\alpha)}^{i}~ D_{i}f~.
\label{stringleib}
\eeq
In this way we may imagine an expression with an arbitrary even 
(non-negative) number of \mb{a_{(\alpha)}}'s.    
For instance, one may check that 
\beq
\int db_{(\alpha)i}~ f~a_{(\alpha)}~g~a_{(\alpha)}~
f'~a_{(\alpha)}~g'~a_{(\alpha)}~f''~=~f f' f''~D_{i}(gg')~.
\eeq
Here we see that the \mb{a}'s plays the role of ``start and stop 
signs'' for the action of the derivatives.
The Leibnitz rule is implemented via the assertion  \eq{stringleib}.
An integration by part of an expression may be performed by cyclicly 
moving an \mb{a_{(\alpha)}} from one end of the expression to the other end 
under an additional change of sign.
Let now \mb{s} and \mb{s'} be two singular expression, cf.\ point \mb{B1} in
Subsection~\ref{secsuitform}.
Some of the consequences of the annihilation principle may now be stated as
\beq
s(x_{(\alpha)})\left[a_{(\alpha)}, ~s'(x_{(\alpha)})\right] ~=~0~,~~~~~~
s(x_{(\alpha)})\left[a_{(\alpha)},~\delta(x_{(\alpha)}\!-\!y) \right] 
s'(y)~=~0~.
\eeq

\end{appendix}




\end{document}